\documentclass[3p,sort&compress,11pt]{elsarticle}
\usepackage{epsf,epsfig,float,latexsym,rotating,graphicx,slashed,bm,ulem}
\usepackage{amsmath,amssymb,amsfonts}
\usepackage[dvipsnames,usenames]{color}

\makeatletter
\newlength{\MySep}
\newcommand{\beq}{\begin{equation}}
\newcommand{\eeq}[1]{\label{#1}\end{equation}}
\newcommand{\beqa}{\begin{eqnarray}}
\newcommand{\eeqa}[1]{\label{#1}\end{eqnarray}}
\newcommand{\eeqan}{\end{eqnarray}}

\newcommand{\CSWT}{$\rm P_{\rm WT}$}
\newcommand{\CSnlo}{$\rm P_{\rm NLO}$}
\newcommand{\GOlo}{$\rm M_{\rm LO}$}
\newcommand{\GOi}{$\rm M_{\rm I}$}
\newcommand{\GOii}{$\rm M_{\rm II}$}
\newcommand{\IHWWT}{$\rm KM_{\rm WT}$}
\newcommand{\IHWlo}{$\rm KM_{\rm LO}$}
\newcommand{\IHWnlo}{$\rm KM_{\rm NLO}$}
\newcommand{\MMii}{$\rm B_{2}$}
\newcommand{\MMiv}{$\rm B_{4}$}

\def\gsim{\compoundrel>\over\sim}

\def\compoundrel#1\over#2{\mathpalette\compoundreL{{#1}\over{#2}}}
\def\compoundreL#1#2{\compoundREL#1#2}
\def\compoundREL#1#2\over#3{\mathrel
      {\vcenter{\hbox{$\m@th\buildrel{#1#2}\over{#1#3}$}}}}

\makeatother
\journal{Nuclear Physics A}

\begin{document}

\begin{frontmatter}

\title{On the pole content of coupled channels chiral approaches\\
used for the $\bar{K}N$ system}

\author[UJF]{A.~Ciepl\'{y}}
\author[BU]{M.~Mai}
\author[BU,J]{Ulf-G.~Mei{\ss}ner}
\author[UTEF]{J.~Smejkal}
\address[UJF]{Nuclear Physics Institute, 250 68 \v{R}e\v{z}, Czech Republic}
\address[BU]{Helmholtz-Institut f\"{u}r Strahlen- und Kernphysik (Theorie) and 
Bethe Center for Theoretical Physics, Universit\"{a}t Bonn, D-53115 Bonn, Germany}
\address[UTEF]{Institute of Experimental and Applied Physics, Czech Technical 
University in Prague, Horsk\'{a} 3a/22, 128~00~Praha~2, Czech Republic}
\address[J]{Institute for Advanced Simulation (IAS-4), Institut f\"ur Kernphysik (IKP-3)
and J\"ulich Center for Hadron Physics,
Forschungszentrum J\"ulich, D-52425 J\"ulich, Germany}

\begin{abstract}
Several theoretical groups describe the antikaon-nucleon interaction at low energies 
within approaches based on the chiral SU(3) dynamics and including next-to-leading 
order contributions. We present a comparative analysis of the pertinent models 
and discuss in detail their pole contents. It is demonstrated 
that the approaches lead to very different predictions for the $K^{-}p$ amplitude 
extrapolated to subthreshold energies as well as for the $K^{-}n$ amplitude. 
The origin of the poles generated by the models is traced to the so-called
zero coupling limit, in which the inter-channel couplings are switched off. 
This provides new insights into the pole contents of the various approaches.
In particular, different concepts of forming the $\Lambda(1405)$ 
resonance are revealed and constraints related to the appearance 
of such poles in a given approach are discussed.
\end{abstract}

\begin{keyword}
chiral dynamics \sep antikaon-nucleon interaction \sep baryon resonances 

\PACS 11.80.Gw \sep 12.39.Fe \sep 12.39.Pn \sep 14.20.Gk
\end{keyword}

\end{frontmatter}

\section{Introduction}
\label{sec:intro}

The modern treatment of the $\bar{K}N$ interactions at low energies is based on Chiral Perturbation 
Theory ($\chi$PT), an effective field theory \cite{Weinberg:1978kz,Gasser:1983yg,Gasser:1984gg} 
that implements the QCD symmetries in the region of the large strong coupling constant. 
Originally,  $\chi$PT was designed to treat the strong interactions 
between mesons, the Goldstone bosons of the theory. Later on, it was extended to the meson-baryon 
sector, see Refs.~\cite{Gasser:1987rb,Krause:1990xc} for initial works. The power counting
issues arising from the large nucleon mass in the chiral limit were overcome by various
methods, for a recent review see Ref.~\cite{Bernard:2007zu}. Using baryon $\chi$PT, 
the $\bar{K}N$ interaction was analyzed  in Ref.~\cite{Kaiser:1995eg}, where it was 
shown that the standard perturbation series does not converge due 
to strong coupling of $\bar{K}N$ to the $\pi\Sigma$ channel and due to an existence of the 
$\Lambda(1405)$ resonance below the $\bar{K}N$ threshold (see also Ref.~\cite{Lee:1994my}).
In Ref.~\cite{Kaiser:1995eg} the problem was overcome with an introduction of effective 
pseudo-potentials 
constructed to match the chiral amplitudes in the Born approximation. Within this 
quantum-mechanical approach, the use of multi-channel techniques and Lippmann-Schwinger 
equation then allows to sum properly a presumably dominant part of the full perturbation series. 
Another way of dealing with the divergences was adopted in \cite{Oset:1997it} where 
the intermediate state Green function was regularized by means of a momentum cutoff. 
The authors of Ref.~\cite{Oset:1997it} also argued in favor of neglecting off-shell effects 
inherent in the separable model of Ref.~\cite{Kaiser:1995eg} and introduced channels closed 
at the $\bar{K}N$ threshold to improve description of experimental data. Finally, the chiral 
approach to $\bar{K}N$ interactions was reformulated within a complete quantum field 
methodology based on dispersion relation for the inverse of the scattering matrix where 
dimensional regularization was used to tame the infinities of the meson-baryon 
Green function~\cite{Oller:2000fj}. In such an approach unitarity is preserved at each order 
of the chiral expansion of the potential \cite{Oller:2000fj,Jido:2003cb}. 
For latter use, the approach is often referred to as the chiral unitary model.
It was also shown how one can match the so-constructed non-perturbative amplitudes to the 
$\chi$PT amplitudes beyond leading order. 

In a broad region around the $\bar{K}N$ threshold the energy and density dependence 
of the isoscalar part of the $\bar{K}N$ amplitude is strongly influenced 
by the $\Lambda(1405)$ resonance. While the $\Lambda(1405)$ properties 
have been well known for a long time, the nature of the resonance retains some 
unresolved mysteries. At present, the $\Lambda(1405)$ appears to be dynamically
generated, see e.g.~the study of compositeness in Ref.~\cite{Kamiya:2016jqc}, 
but the situation is complicated by the appearance of two close-by poles.  
The reader is also referred to the in-depth review by Hyodo and Jido \cite{Hyodo:2011ur} as 
well as to a very recent review in the PDG tables~\cite{PDGreview}.

The chiral approaches that implement the strong coupling of the $\pi \Sigma$ and $\bar{K}N$ channels 
lead to two dynamically generated resonances assigned to the $\Lambda(1405)$, each of them 
coupling individually to the $\bar{K}N$ and $\pi\Sigma$ states \cite{Oller:2000fj,Jido:2003cb}.
Although the theoretical models give different predictions for 
exact positions of the poles it seems that the very recent experimental measurements 
of the kaonic hydrogen characteristics  by the SIDDHARTA collaboration \cite{Bazzi:2011zj} 
based on the improved Deser-type formula \cite{Meissner:2004jr}
combined with an analysis of the $\pi \Sigma$ mass spectra 
observed in photoproduction experiments by the CLAS collaboration \cite{Moriya:2013eb} help 
to fix at least the position of the pole at higher energy that couples more strongly to the $\bar{K}N$ channel.
Particularly, the CLAS data may be used to choose among a number of local minima emerging 
in $\chi ^{2}$ fits with many free parameters introduced when next-to-leading order (NLO) 
corrections are accounted for in the meson-baryon interactions \cite{Mai:2014xna}.

In the present work we look at another aspect of the theory, the origin of the poles 
generated by the multi-channel dynamics. It was already shown by Hyodo and Weise \cite{Hyodo:2007jq} 
that one of the $\Lambda(1405)$ related states transforms into a $\pi\Sigma$ resonance 
and the other one into a $\bar{K}N$ bound state when the inter-channel couplings 
are switched off. In the current work we elaborate on this finding within a more general 
concept, discuss the conditions under which such poles emerge and display the pole 
content of several different models, all of them reproducing the same experimental data.
By requiring an existence of a pole in a particular channel we get additional restrictions 
on the values of some parameters that are fitted to the experimental data.  
We also maintain that only a limited number of poles can be generated dynamically within 
the coupled channels framework with interactions driven by chiral symmetry. 

The paper is outlined as follows. In the next section we briefly review the chiral unitary 
approach to meson-baryon interactions and specify the models used in our analysis. Then, 
we compare the predictions of the models for the energy dependence of the $\bar{K}N$ 
amplitude as well as for the isoscalar poles assigned to the $\Lambda$(1405) resonance. 
Further, in Section~\ref{sec:poles} we derive conditions on pole existence in a limit 
of switched-off inter-channel couplings and continue with a discussion of the pole 
contents of the selected models. Finally, we conclude the paper with a summary.

\section{Coupled channels approaches}
\label{sec:models}

In our analysis we focus on $\bar{K}N$ interactions described within a framework 
of coupled channels approaches with inter-channel couplings derived from an  
effective chiral Lagrangian. The channels involve all meson-baryon states built 
up from the corresponding 
ground state octets with the total strangeness $S=-1$. In order of the corresponding 
production threshold energies these are: $\pi\Lambda$, $\pi\Sigma$, $\bar{K}N$, 
$\eta\Lambda$, $\eta\Sigma$ and $K\Xi$, with appropriate charges (or isospins). 
In terms of threshold energies these channels encompass a broad interval from 1250~MeV 
to about 1800~MeV. However, it should be noted that we restrict ourselves to two-body 
channels in $s$-wave only, so that any effects due to a formation of additional particles are neglected 
in the approaches discussed here. Additionally, since the model parameters are fitted to experimental 
data available at the $K^{-}p$ threshold and at low kaon momenta it is clear that the models 
may lack a predictive power at energies far from the $\bar{K}N$ threshold. 
However, for the study of the qualitative features related to dynamically generated 
resonances and to the SU(3) symmetry of the chiral interactions such approach is perfectly suited.

In the following we will outline the theoretical approaches used in the present work. 
We will concentrate on the most important features of these, referring the interested 
reader to the original publications for more details. All models considered here 
describe (at least) the available total cross sections 
$K^-p\to K^-p,~\bar K^0 n,~\pi^0\Sigma^0,~\pi^+\Sigma^-,~\pi^-\Sigma^+$ at low energies, 
see Refs.~\cite{Ciborowski:1982et, Humphrey:1962zz, Sakitt:1965kh,  Watson:1963zz}, 
the old but very precise threshold branching ratios 
\begin{align}\label{eq:BRs}
&\gamma=\frac{\Gamma(K^-p\rightarrow
 \pi^+\Sigma^-)}{\Gamma(K^-p\rightarrow \pi^-\Sigma^+)}\,,~
 R_n=\frac{\Gamma(K^-p\rightarrow
 \pi^0\Lambda)}{\Gamma(K^-p\rightarrow 
 \text{neutral states})}\,,~
 R_c=\frac{\Gamma(K^-p\rightarrow \pi^+\Sigma^-, \pi^-\Sigma^+)}{\Gamma(K^-p\rightarrow \text{inelastic channels})}\,
\end{align}
from Refs.~\cite{Tovee:1971ga,Nowak:1978au}, as well as the energy shift and width due 
to strong interaction  measured for the 1s level of the kaonic hydrogen in the SIDDHARTA experiment at DA$\Phi$NE \cite{Bazzi:2011zj}. 
Besides that, some of the considered models included additional experimental data to constrain 
a relatively large parameter space as we specify below.

On the theoretical side, the starting point of all considered approaches is the meson-baryon 
potential derived from the chiral Lagrangian, which at the leading chiral order reads
\begin{align}\label{eq:LAGRlo}
   \mathcal{L}^{(1)}_{\phi B}=\langle \bar{B} (i\gamma_\mu D^\mu-M_0)B\rangle
   +\frac{D}{2}\langle \bar{B}\gamma_\mu \gamma_5[u^\mu,B]_+ \rangle
   +\frac{F}{2}\langle \bar{B}\gamma_\mu \gamma_5[u^\mu,B] \rangle \,.
\end{align}
Here $\langle\ldots\rangle$ denotes the trace in the flavor space, $D_\mu B:=\partial_\mu B 
+\frac{1}{2}[[u^\dagger,\partial_\mu u],B]$, and $D,\,F$ are the axial coupling constants. 
The traceless meson matrix $\phi$, included in the above Lagrangian via $u^2:=
\exp\bigl(i\phi/F_0\bigr)$ and $u^\mu:=iu^{\dagger}\partial^\mu u  - iu\partial^\mu u^{\dagger}$, 
collects the Goldstone bosons of the theory, i.e. $\pi, K,\eta$. The ground state octet 
of baryons ($N, \Sigma, \Lambda, \Xi$) is included via the traceless matrix $B$. All external 
currents are set to zero except the scalar one, which reflects the explicit chiral 
symmetry breaking and is set equal to the quark mass matrix $\mathcal{M}:=\textrm{diag}(m_u,m_d,m_s)$. 
Note, however, that in $\chi$PT the quark masses always appear in combination with the low-energy
constant (LEC) $B_0$ that measures the strength of the scalar quark condensate.
Finally, the meson decay constant $F_0$ as well as the baryon octet mass $M_0$ are the 
values in the chiral $SU(3)$ limit. In the models discussed in our work, they are set to 
their physical values in the pertinent channels. At the leading order (LO) 
the meson-baryon interaction is given by the so-called Weinberg-Tomozawa (WT) term derived 
from the covariant derivative $D_\mu B$. Moreover, the axial vector current $\sim D$, $F$ gives 
rise to the so-called Born graphs, which encounter for the $s$- and $u$-channel exchange diagrams 
of the intermediate baryons.

At the next-to-leading chiral order the meson-baryon interaction is given entirely by 
contact terms. The details on the construction of the NLO Lagrangian as well as its full 
form can be found in Refs.~\cite{Krause:1990xc,Frink:2004ic}.
The LECs that are introduced at this order encompass three so-called 
{\it symmetry breakers}  (responsible for the splitting of baryon masses) as well as eleven 
additional LECs commonly referred to as  {\it dynamical} LECs. Note, that putting the intermediate 
particles on their mass shell and projecting to the $s$-wave
the number of dynamical structures can be reduced to four. 
For our purposes, here we only remark that 
the symmetry  breakers can in principle be related to the baryon masses and to the pion-nucleon
$\sigma$ term while the dynamical 
LECs are very loosely restricted. Typically, they are treated as free  parameters. 
 
In the paper we discuss all currently available theoretical approaches that incorporate NLO terms 
in the underlying chiral Lagrangian and determine the free model parameters by $\chi^{2}$ 
fits that reproduce (among others) the most recent SIDDHARTA data on kaonic hydrogen.  
Before doing so, one important remark is in order. According to the power counting, one expects the
contribution of the NLO terms to be parametrically suppressed compared to the LO one. However, there
can be exceptions to this. It is known that in chiral SU(3) one can have large corrections due to kaon loops.
However, as argued in Ref.~\cite{Jenkins:1991bs}, the first correction can be large
and after that, the series behaves as expected. Such a behavior is indeed observed in the chiral expansion
of the baryon octet magnetic moments~\cite{Meissner:1997hn}. Another possible enhancement of formerly
subleading terms can arise from the presence of close-by resonances, as it is well known from the chiral
expansion of the two-nucleon forces. Here, the N$^2$LO corrections to some phase shifts are as large
as the NLO ones, as the effects of the close-by $\Delta$ are encoded in the LECs $c_i$, which contribute
first to the first corrections of the two-pion exchange, that appear at N$^2$LO, see e.g. the review~\cite{Epelbaum:2008ga}.
Therefore, in some of the approaches discussed below, the formal suppression of the NLO terms is not imposed.

~\\\noindent
\textbf{Model types: Kyoto-Munich, Murcia}

The Kyoto-Munich and Murcia approaches for the antikaon-nucleon scattering rely 
on a similar theoretical assumption as described in Refs.~\cite{Ikeda:2012au} 
and~\cite{Guo:2012vv}, respectively. 
There, the chiral potential of the leading and next-to-leading order, derived as described 
above, is put on the mass shell and projected onto the $s$-wave. The exact form of the 
projected chiral potentials and corresponding coupling matrices dictated by the SU(3) symmetry 
can be found in e.g. Ref.~\cite{Borasoy:2005ie}. In the next step, the scattering amplitude 
$T(\sqrt{s})$ is derived, resumming this potential in the coupled-channel Bethe-Salpeter 
equation in the on-shell approximation\footnote{Originally, this form of the scattering 
amplitude was derived taking into account the unitarity cut via the dispersion relations 
in Ref.~\cite{Oller:2000fj}.}
\begin{align}\label{eq:BSEonshell}
 T(\sqrt{s})=V(\sqrt{s})+V(\sqrt{s})\:G(\sqrt{s})\:T(\sqrt{s})\,,
\end{align}
where every element represents a matrix in the channel space. The diagonal matrix 
$G(\sqrt{s})$ contains the logarithmically (UV) divergent one-loop meson baryon integrals, 
which are tamed by dimensional regularization applying the $\overline{\rm MS}$ subtraction 
scheme. In four dimensions and for a specific meson-baryon channel index $i$ it reads
\begin{align}\label{eq:MBloop}
G_i(\sqrt{s})=\frac{1}{16 \pi^2} \left[ a_i(\mu) + \log \frac{M_i^2}{\mu^2} \right] 
+ G_{0}(\sqrt{s},m_i,M_i)\,,
\end{align}
where $M_i (m_i)$ denotes the baryon (meson) masses of the intermediate particles and we introduced
\begin{align}\label{eq:G0}
G_{0}(\sqrt{s},m_i,M_i) = - \frac{1}{16 \pi^2} 
                          + \frac{m_i^2- M_i^2 + s}{32\pi^2 s}\log\left(\frac{m_i^2}{M_i^2}\right)
                          - \frac{q_i(\sqrt{s})}{4\pi^2 \sqrt{s}}\mathrm{artanh}\left(\frac{2q_i(\sqrt{s}) \sqrt{s}}{(M_i + m_i)^2 - s}\right)
\end{align}
for the scale-independent part of the loop function, with $\mu$ the scale of dimensional
regularization and $q_i(\sqrt{s})$ the modulus of the center-of-mass (CMS) 
three momentum in the pertinent channel. At any order of a full perturbative 
calculation this scale dependence cancels out the scale dependence of the contact terms. 
In a non-perturbative analysis relying on a driving term of any fixed chiral order 
the scale dependence remains, reflecting the missing higher order (local) terms not accounted for 
in the above definition of the scattering matrix. Note also that the higher order corrections 
are channel dependent. Therefore, to account for the missing topologies, the subtraction 
constants $a_{i}(\mu)$ are used in each interaction channel as free parameters of the model. 
They are adjusted together with the low-energy constants to reproduce the experimental data.
The exact values of the subtraction constants depend on the regularization scale $\mu$ adopted to be the 
same for all channels. The Kyoto-Munich \cite{Ikeda:2012au} and Murcia \cite{Guo:2012vv} approaches 
use the subtraction constants as free parameters for $\mu=1$~GeV and $\mu=0.77$~GeV, respectively.

The free parameters are adjusted in both approaches to reproduce the experimental data specified before. 
Additionally, in the Murcia approach cross sections on $K^-p\to \eta\Lambda$, $K^-p\to \pi^0\pi^0\Sigma^0$ 
as well as the line shape of $K^-p\to \pi^0\pi^0\Sigma^0$ from 
Refs.~\cite{Prakhov:2004an,Starostin:2001zz,Hemingway:1984pz} are used to constrain the parameter 
space. Furthermore, $\sigma_{\pi N}$, $a_{0+}^{+}$ and $N$, $\Lambda$, $\Sigma$, $\Xi$ 
masses are used to restrict the unknown symmetry breakers, i.e. $b_0$, $b_D$, $b_F$. 
Various scenarios are analyzed in both papers, regarding the form 
of the chiral potential $V(\sqrt{s})$. In Ref.~\cite{Ikeda:2012au} fits are performed 
for the potential containing only WT terms (\IHWWT), all terms of the first chiral 
order (\IHWlo), as well as all terms of first and second chiral order (\IHWnlo). 
In Ref.~\cite{Guo:2012vv} fit results are presented for the chiral potential 
of the leading (\GOlo) as well as for the full chiral potential of the second 
chiral order (\GOi{} and \GOii).

~\\\noindent
\textbf{Model type: Bonn}

The Bonn approach is used to describe the aforementioned experimental data 
in Ref.~\cite{Mai:2014xna}. Derived originally in Refs.~\cite{Mai:2012dt,Bruns:2010sv,Mai:2013cka}, 
it starts from the leading and next-to-leading order chiral Lagrangian.
Without making use of the on-shell approximation the exactly unitary scattering matrix 
$\tilde T(\sqrt{s})$ can be derived for any given kernel $\tilde V$ via the Bethe-Salpeter equation
\begin{align}\label{eq:BSEfull}
\tilde T(\slashed{q}_2, \slashed{q}_1; p)&=\tilde V(\slashed{q}_2, \slashed{q}_1;p) 
      +\int\frac{d^d l}{(2\pi)^d}\tilde V(\slashed{q}_2, \slashed{l}; p) 
      \frac{i}{\slashed{p}-M+i\epsilon}
      \frac{1}{k^2-m^2+i\epsilon}
      \tilde T(\slashed{l}, \slashed{q}_1; p)\,,
\end{align}
where $p$, $q_1$ and $q_2$ denote the overall, incoming and outgoing meson four-momenta, respectively. 
Furthermore, every component of the above integral equation is a $10\times10$ matrix in 
the channel space. Considering local terms only, the solution of the above coupled-channel 
integral equation can be derived exactly, see chapter~2 of Ref.~\cite{Mai:2013cka}. In its 
final form it contains  scalar one-meson(baryon) as well as one-meson-one-baryon loop integrals 
only, treated by dimensional regularization and applying the $\overline{\rm MS}$ subtraction 
scheme. 
There, the regularization scales $\mu_i$ are set to be channel dependent, replacing
\begin{align}\label{eq:mui}
\log \frac{M_i^2}{\mu^{2}_{i}} = a_i(\mu) + \log \frac{M_i^2}{\mu^2}
\end{align}
in Eq.~\eqref{eq:MBloop} to account for the missing (channel dependent) higher order terms.
For the reasons given in Ref.~\cite{Bruns:2010sv} the purely 
baryonic integrals are set to zero from the beginning, whereas the pure mesonic scalar loop 
integrals can be shown to parametrize the off-shell part of the above integral equation. 
The quantitative influence of the latter on the antikaon-nucleon amplitude has been studied 
in Ref.~\cite{Mai:2012dt}, where it was found to be negligible. Thus, in 
 the model of Ref.~\cite{Mai:2014xna} they have been set to zero\footnote{Since no partial 
 wave expansion of the interaction kernel is performed here, 
the scattering matrix $\tilde T$ does not have to coincide with the one from 
Eq.~\eqref{eq:BSEonshell}. It does so, if the kernel contains only $s$-wave contributions, 
e.g.~WT term, which we have checked explicitly.}, which allows to increase the computational 
performance by a factor of~~$\sim 30$. Note that the computational improvement is essential 
for the extensive study of the high-dimensional parameter space (20), given by the 
unknown LECs (14) and regularization scales $\mu_i$ (6). After the best parameter set is found 
one can gradually turn on the off-shell part again, re-adjusting the parameters to fit the 
experimental data.

In a large scale study of the parameter space it has been found in Ref.~\cite{Mai:2014xna} that 
at least 8 different sets of parameters lead to the similar description of the experimental (hadronic)
data. One can,  however, put more constraints on these solutions considering the recent and very 
precise photoproduction data, measured by the CLAS collaboration~\cite{Moriya:2013eb}. Thus, 
originally 8 solutions have been restricted to two solutions to be referred later as the
models \MMii{} and \MMiv{}, respectively, which are the solution~$\#2$ and solution~$\#4$ 
from Ref.~\cite{Mai:2014xna}, correspondingly.

~\\\noindent
\textbf{Model type: Prague}

In the {\it on-shell models} described so far the intermediate state Green function is dimensionally regularized,
see Eq.~(\ref{eq:MBloop}). Alternatively, one can introduce 
a momentum cutoff as it was done in Ref.~\cite{Oset:1997it}. Another way of taming the undesired 
high momenta in the Green function 
relies on adopting the meson-baryon effective potentials in a separable form, 
\begin{align}\label{eq:Vsep}
V_{ij}(k,k';\sqrt{s}) =  g_{i}(k^{2}) \: v_{ij}(\sqrt{s}) \: g_{j}(k'^{2})
\end{align}
with the off-shell form factors chosen in the Yamaguchi form, which reads
\begin{align}\label{eq:FF}
g_{j}(k)=\frac{1}{1+(k/ \alpha_{j})^2} \,.
\end{align}
The inverse ranges $\alpha_j$ characterize the interaction range of the specific 
meson-baryon states. They can be related to the subtraction constants of Eq.~(\ref{eq:MBloop}), 
see Ref.~\cite{Oller:2000fj}, and their exact values are to be determined by a fit to 
experimental data. The effective separable potentials were applied to the $\bar{K}N$ interaction 
for the first time in Ref.~\cite{Kaiser:1995eg} where the authors considered 
only channels that are open at the $\bar{K}N$ threshold. 
The model was extended to cover all interactions of the lightest meson octet 
with the lightest baryon octet in \cite{Cieply:2009ea} and \cite{Cieply:2011nq}, where the parameters 
of the model were also adjusted to reproduce the most recent data on the 1s level 
characteristics in kaonic hydrogen \cite{Bazzi:2011zj}. 

The central piece of the chirally motivated potential matrix $V_{ij}(k,k';\sqrt{s})$ 
in Eq.~\eqref{eq:Vsep} reads as
\begin{align}\label{eq:vij}
v_{ij}(\sqrt{s}) = -\frac{C_{ij}(\sqrt{s})}{4\pi f_{i}f_{j}}\: \sqrt{\frac{M_i M_j}{s}}\,,           
\end{align}
which has a form that matches the chiral amplitude up to a given order in the meson momenta. 
Here again, $f_{j}$ denotes the meson decay constant and $M_{j}$ the baryon mass 
in the $j$-th channel. The couplings $C_{ij}$ are determined by the SU(3) symmetry and 
include energy-dependent contributions derived from the chiral Lagrangian in the same 
manner as it is done for the {\it on-shell models} Kyoto-Munich and Murcia. In Ref.~\cite{Cieply:2011nq} 
two approaches were presented, one with the potential matrix restricted to only WT interactions 
and the second including all NLO terms. In the present work we will refer to these 
Prague models as to \CSWT{} (called TW1 in Ref.~\cite{Cieply:2011nq}) 
and \CSnlo{} (NLO30 in Ref.~\cite{Cieply:2011nq}), respectively. 
We also note that in the \CSnlo{} model the inverse ranges of the channels 
closed at the $\bar{K}N$ threshold as well as the symmetry breakers $b_0$, $b_F$ 
and $b_D$ were fixed in the fits to reduce the number of the free parameters 
to $7$, the lowest among the considered NLO approaches. 

With the potential kernel of Eq.~(\ref{eq:vij}) the loop series (Lippmann-Schwinger equation) 
can be solved exactly to obtain the meson-baryon amplitudes in a separable form as well,
\begin{align}\label{eq:Fsep}
F_{ij}(k,k';\sqrt{s}) = g_{i}(k^{2}) \: f_{ij}(\sqrt{s}) \: g_{j}(k'^{2}) \,.
\end{align}
The algebraic solution for the reduced (stripped off the form factors) amplitudes then 
reads
\begin{align}\label{eq:fij}
f_{ij}(\sqrt{s}) = \left[ (1 - v \cdot G(\sqrt{s}))^{-1} \cdot v \right]_{ij}
\end{align}
with an intermediate state meson-baryon Green function
\begin{align}\label{eq:Gn}
G_{n}(\sqrt{s}) & =  -4\pi \: \int\frac{d^{3}p}{(2\pi)^{3}}\frac{g_{n}^{2}(p^{2})}{k_{n}^{2}-p^{2} +i\epsilon} 
                 =  \frac{(\alpha_n + ik_n)^{2}}{2\alpha_{n}}\:[g_{n}(k_{n})]^{2} \,.
\end{align}
The resulting separable amplitudes provide a natural tool for testing the meson-baryon 
interactions off the energy shell, particularly in nuclear matter \cite{Cieply:2011yz, Cieply:2011fy,Cieply:2015pwa}.

\vspace*{5mm}
We close this section with a brief comparison of the approaches by looking at the impact of including 
the NLO terms on the inter-channel couplings. For this purpose we write the coupling matrix $C_{ij}$ 
in a form
\begin{align}\label{eq:CWT}
C_{ij}(\sqrt{s}) = -\tilde{C}_{ij}(\sqrt{s}) (2\sqrt{s} - M_i -M_j)/4  \,,
\end{align}
where the tilded renormalized matrix $\tilde{C}_{ij}(\sqrt{s})$ is introduced in a way to match
the standard energy independent SU(3) Clebsch-Gordan coefficients if only the WT interaction 
is considered. To simplify the discussion we concentrate only on the diagonal 
couplings evaluated at the pertinent channel thresholds. In the Table~\ref{tab:CNLO} we show 
the isoscalar ($I=0$) and isovector ($I=1$) diagonal renormalized couplings 
$\tilde{C}_{ii}(\sqrt{s}=M_{i}+m_{i})$ calculated for the considered models.
Note that the Bonn model does not introduce a concept of a potential. 
In order to perform the proposed comparison of the threshold couplings 
we have calculated the $s$-wave part of the $T$-matrix from Eq.~\eqref{eq:BSEfull} 
with one-meson-one-baryon loops set to zero. In each channel the result 
was normalized such that it reproduces WT values (first line of Table~\ref{tab:CNLO}) 
exactly, when the NLO terms are set to zero. We also remark that the free parameters 
of the NLO potential and those of the loop functions are fitted simultaneously, 
so the strength can be reshuffled between the LECs and the subtraction constants. 
For this reason the deviations of the $\tilde{C}_{ij}$ couplings from the WT ones 
serve only as an indirect indication of how large the NLO contributions are.

\begin{table}[htb]
\caption{The diagonal values of the inter-channel couplings normalized to match the SU(3) 
Clebsch-Gordan coefficients of the WT interaction when other {\it non WT} contributions are set 
to zero. The couplings are evaluated at the pertinent channel thresholds in the isospin basis.
}
\begin{center}
\begin{tabular}{|c|cccc|ccccc|}
\hline
          &\multicolumn{4}{c|}{$I=0$}& \multicolumn{5}{c|}{$I=1$} \\
\hline
 model    & $\pi\Sigma$ & $\bar{K}N$ & $\eta\Lambda$ & $K\Xi$ & $\pi\Lambda$ & $\pi\Sigma$ & $\bar{K}N$ & $\eta\Sigma$  & $K\Xi$ \\
\hline
  WT      &   $4.00$    &   $3.00$   &    $0.00$     & $3.00$ &    $0.00$    &   $2.00$    &   $1.00$   &    $0.00$    & $1.00$ \\
\IHWnlo{} &   $4.50$    &   $2.93$   &    $0.66$     & $3.04$ &    $0.07$    &   $1.68$    &   $1.11$   &    $0.30$    & $1.05$ \\
\CSnlo{}  &   $3.98$    &   $2.79$   &   $-0.15$     & $3.42$ &   $-0.06$    &   $2.00$    &   $1.14$   &   $-0.16$    & $0.51$ \\
\GOi{}    &   $7.48$    &  $11.36$   &    $8.79$    & $13.67$ &    $0.21$    &   $2.15$    &   $3.48$   &    $0.86$    & $1.22$ \\
\GOii{}   &   $6.45$    &   $8.60$   &    $5.48$     & $9.30$ &   $-0.01$    &   $1.79$    &   $1.63$   &    $0.17$    & $0.97$ \\
\MMii{}   &   $4.22$    &  $18.49$   &   $-0.59$     & $4.77$ &   $-1.22$    &   $3.26$    &  $-8.37$   &   $-3.48$    & $5.20$ \\
\MMiv{}   &   $3.72$    &   $6.47$   &   $-1.56$    & $-4.35$ &    $1.05$    &   $2.59$    &  $-1.51$   &    $0.70$    & $6.39$ \\
\hline
\end{tabular}
\end{center}
\label{tab:CNLO}
\end{table}

The \IHWnlo{} and \CSnlo{} models introduce only moderate corrections to the couplings 
when compared with the WT values shown in the first row of Table \ref{tab:CNLO}. On the other hand, 
the table reveals that the Murcia and Bonn models include quite sizable NLO contributions. 
Particularly, both Murcia models have very large diagonal couplings that deviate significantly 
from the WT values in the whole isoscalar sector. The Bonn models also provide quite large 
$\bar{K}N(I=0)$ couplings, though for the remaining isoscalar channels the NLO modifications 
are not so significant as those in the Murcia models. It is very interesting that the experimental 
data do not exclude large NLO contributions in spite of their good reproduction with 
models implementing only the LO interactions. Since the isovector sector is practically not restricted 
by the experimental data the differences among the models and large NLO contributions in some 
of them are not so surprising there.

\section{Model predictions}
\label{sec:compar}

Since the models are based on the same chiral Lagrangian and reproduce very similar sets 
of experimental data one would assume that they should tend to agree on predictions made 
for other physically relevant quantities like the energy dependence of the $\bar{K}N$ 
amplitudes or the positions of the poles of the amplitudes on various Riemann sheets (RSs) of 
the complex energy manifold. As we will demonstrate, the reality is quite a bit different.

\begin{figure}[thb]
\resizebox{0.5\textwidth}{!}
{\includegraphics{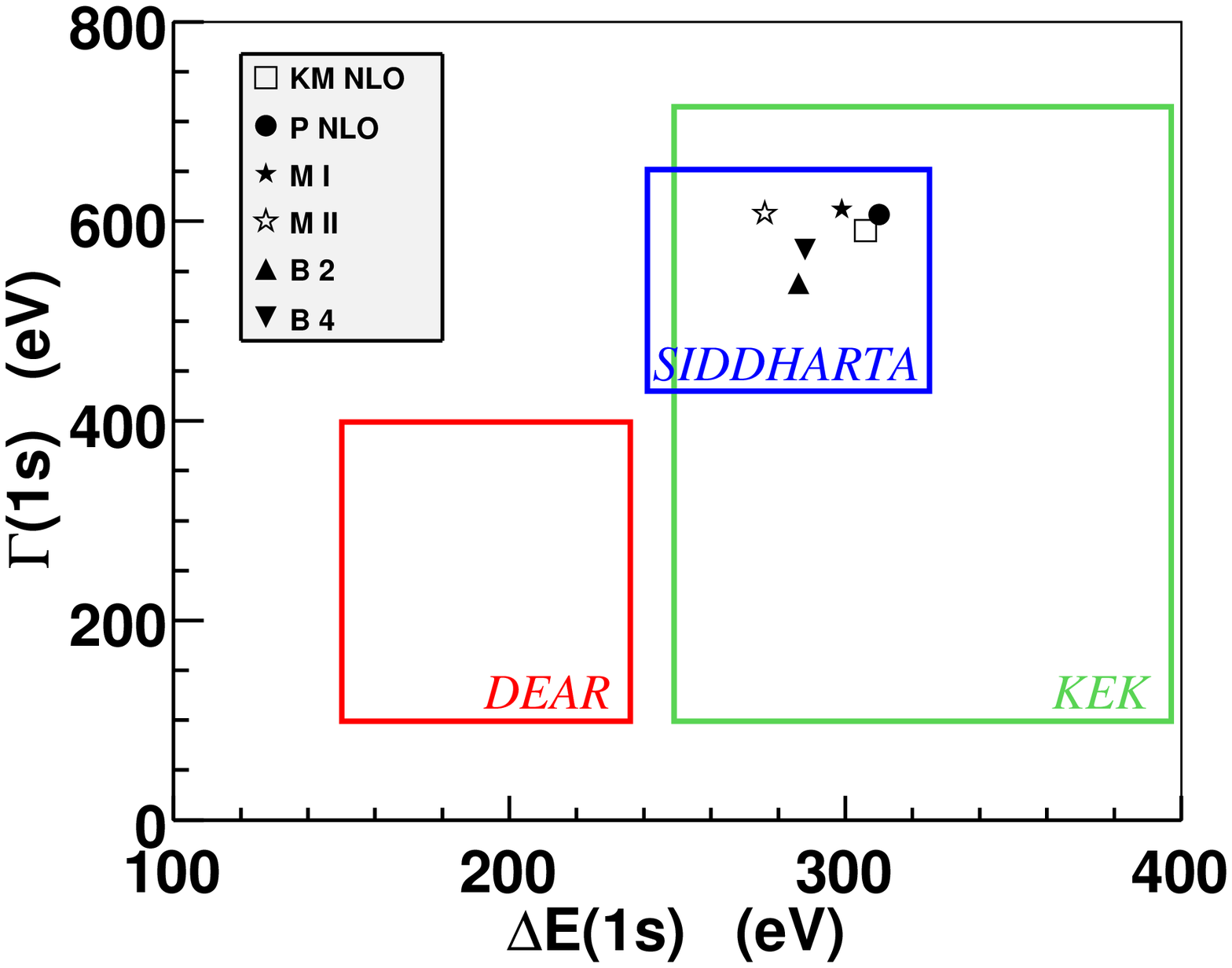}
}
\resizebox{0.5\textwidth}{!}
{\includegraphics{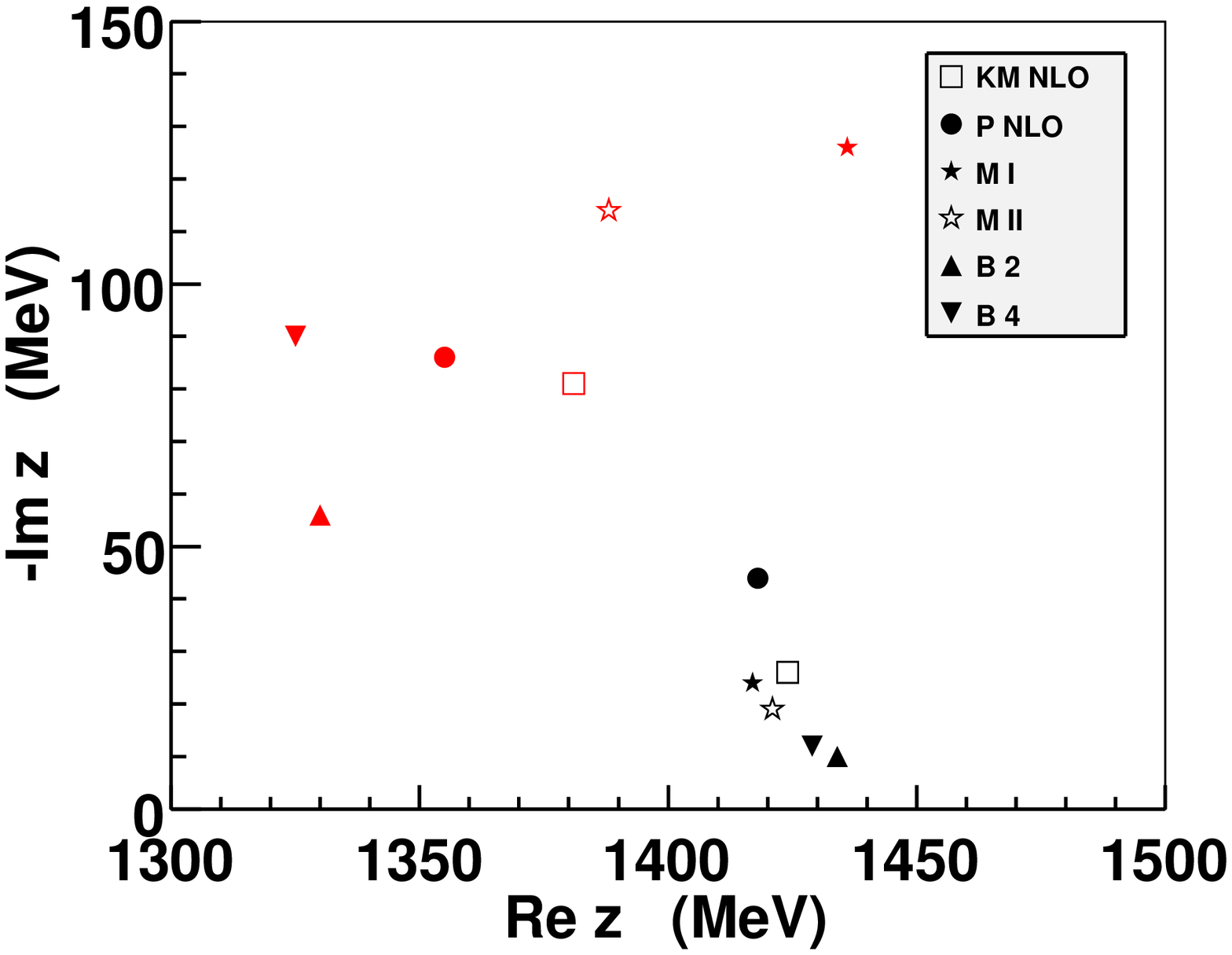}
}
\caption{Kaonic hydrogen characteristics and pole positions for various approaches.}
\label{fig:poles}       
\end{figure}

In the left panel of Figure~\ref{fig:poles} we show how the models reproduce the experimental 
data for the 1s level characteristics of kaonic hydrogen, the energy shift $\Delta E(1s)$ and 
the absorption width $\Gamma (1s)$, both caused by the strong interaction. There, one can also 
see the progress made from the experimental point of view by comparing the oldest 
KEK results \cite{Iwasaki:1997wf} with  those by the DEAR collaboration \cite{Iwasaki:1997wf} 
and the most recent and most precise ones provided by the SIDDHARTA measurement \cite{Bazzi:2011zj}. 
All considered theoretical approaches reproduce 
the SIDDHARTA data quite well and are in very close agreement, even more precise than indicated by 
the experimental standard deviation. The agreement is spoiled when it comes to positions 
of the two poles assigned to the $\Lambda(1405)$ resonance and visualized in the right panel 
of Figure~\ref{fig:poles}. There, the models only agree on the real part of the complex energy 
for the pole that couples more strongly to the $\bar{K}N$ channel and is generated at higher 
energy of about 1420~MeV. The imaginary part of the pole energy is not established so well 
and the position of the second pole varies from one model to another, apparently not constrained 
much by the experimental data. There, the new CLAS data on $\pi\Sigma$ photoproduction off proton 
\cite{Moriya:2013eb} are indeed helpful and were already used in fits to separate between 
many $\chi^{2}$ local minima with the results preferring the models \MMii{} and 
\MMiv{} \cite{Mai:2014xna}. 
On the other hand, the theoretical models still find it difficult to explain the 
peaks in the $\pi\Sigma$ 
mass spectra observed in the $pp$ collisions by the HADES experiment \cite{Agakishiev:2012xk}. 
Apparently, more dedicated studies accounting properly for the dynamics of the $\pi\Sigma$ 
production in those processes are needed before more conclusive results can be reached, 
particularly for the $\pi\Sigma$ related pole.

\begin{figure}[thb]
\resizebox{1.0\textwidth}{!}
{\includegraphics{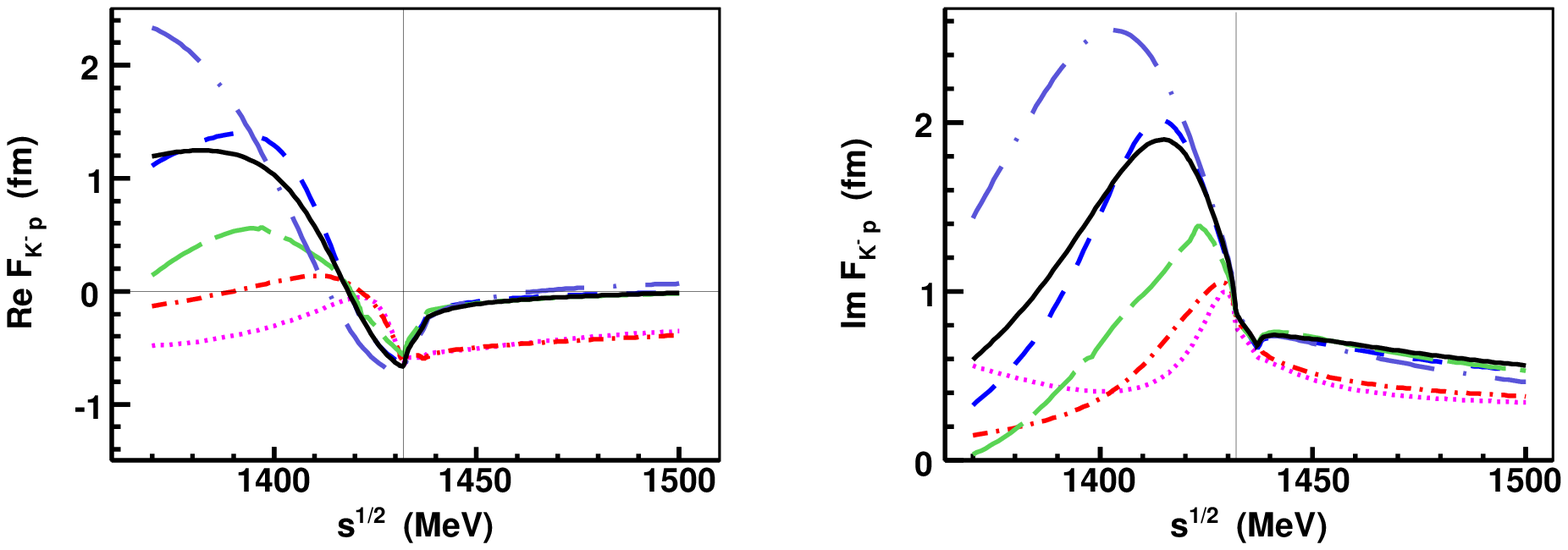}
}\\
\resizebox{1.0\textwidth}{!}
{\includegraphics{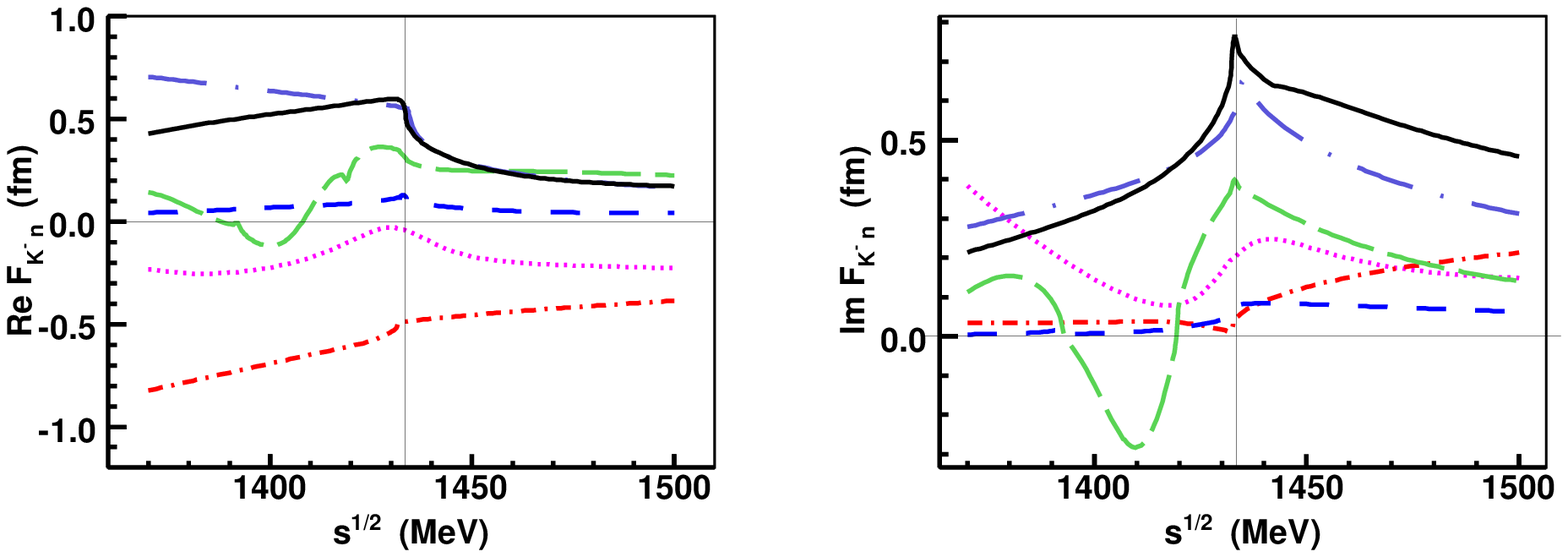}
}
\caption{The $K^{-}p$ (top panels) and $K^{-}n$ (bottom panels) elastic scattering amplitudes 
generated by the NLO approaches considered in our work. The various lines refer to the models:  
\MMii{} (dotted, purple), \MMiv{} (dot-dashed, red), \GOi{} (dashed, blue), 
\GOii{} (long-dashed, green), \CSnlo{} (dot-long-dashed, violet), \IHWnlo (continuous, black).
}
\label{fig:KNampl}       
\end{figure}

The energy dependences of the $K^{-}p$ and $K^{-}n$ amplitudes generated by the NLO models 
are shown in Figure \ref{fig:KNampl}. Once again, the models have no problem to reproduce 
the experimental data available at the $K^{-}p$ threshold energy, the branching ratios 
Eq.~(\ref{eq:BRs}) and the kaonic hydrogen characteristics. However, the models differ 
significantly when going to subthreshold energies in the $K^{-}p$ amplitude or when 
the predictions for the $K^{-}n$ amplitude are made. Particularly, we note that 
the model differences in the subthreshold region are much larger than bands (or zones)
of theoretical uncertainties derived from standard deviations of the fitted parameters 
within a particular approach. Once again, the experimental data available at the threshold 
and at higher energies do not constrain the theoretical models sufficiently 
when exploring physics in those sectors. Thus, further improvement of experimental data, 
e.g.~on the very old cross sections as proposed in Ref.~\cite{Briscoe:2015qia}, 
is of huge importance for this strongly debated field. 
Additional constrains on the $\bar{K}N$ energy dependence at subthreshold energies
should be provided by analysis of the $\pi\Sigma$ mass spectra observed in various 
processes. The CLAS data on $\pi\Sigma$ photoproduction \cite{Moriya:2013eb} were already used 
in Ref.~\cite{Mai:2014xna} while the coming data on the $K^{-}d \rightarrow \pi\Sigma n$ 
reaction and the $\Lambda_{c}\rightarrow \pi\pi\Sigma$ decay can provide 
further constrains on the theoretical models as discussed in Ref.~\cite{Kamiya:2016jqc}. 
However, we note that a proper analysis of the $\pi\Sigma$ spectra observed 
in these processes hinges on realistic treatments of the involved reaction dynamics.

For the $K^{-}p$  amplitude at subthreshold energies the Prague model provides the largest 
and most pronounced attraction (in the real part of the $K^{-}p$ amplitude) and peak 
(in the imaginary part) at energies about 30~MeV below the $\bar{K}N$ threshold with 
a shape in agreement with the Kyoto-Munich and \GOi{} models that have the resonance structure 
at slightly higher energies. The \GOii{} and both Bonn models generate apparently different 
energy dependence, with the Bonn models deviating from the other ones at energies above 
the $\bar{K}N$ threshold as well. It is interesting that both Murcia models agree with 
the Prague and Kyoto-Munich ones at energies above the $\bar{K}N$ threshold despite the fact 
that the Murcia parameter sets were fitted to additional experimental data available 
for higher energies.  

The deviations between the Bonn and other approaches are of great conceptual importance, 
which become most evident at energies above the $\bar{K}N$ threshold. 
They arise due to the fact that, in contrast to the other considered approaches, 
no $s$-wave projection of the interaction kernel is performed in the Bonn approach. 
Therefore, terms of the NLO Lagrangian proportional to the cosine of the 
scattering angle are explicitly accounted for in this approach.
This issue is completely irrelevant when $s$-wave quantities are compared 
(such as scattering lengths), but may play an important role when addressing total 
cross sections. In general, the latter do not discriminate between various partial waves. 
Thus, neglecting higher partial waves in the calculation of the total cross sections 
is an additional assumption. It is not clear {\it a priori}, where this assumption breaks 
down, but it certainly will happen at some energies above the $\bar{K}N$ threshold. 
As a matter of fact, the experimental data in this region are actually dominated 
by the measurement of the total cross sections.

The model predictions for the $K^{-}n$ amplitude were calculated as an isovector $\bar{K}N$ 
amplitude with the hadron masses set to isospin averaged values. The results are shown in the bottom 
panels of Figure \ref{fig:KNampl}. One can see that only the Prague and Kyoto-Munich models are in 
a reasonable agreement here while the other models deviate not only in terms of the amplitude 
magnitude but in the qualitative shape of the energy dependence as well. It is hard to say 
what causes such large deviations, specifically even a change of sign observed for the imaginary 
part of the $K^{-}n$ amplitude for the \GOii{} model. We believe that this issue could be 
resolved if there were available any experimental data on low energy $K^{-}$ scattering off 
the deuterium target as it is hard to foresee a direct measurement of the $K^{-}n$ scattering. 
Dedicated experiments in this direction were proposed in J-PARC \cite{JPARK} and Frascati 
\cite{LNF}. Using Faddeev equations~\cite{Shevchenko:2014uva} or effective field theory 
methods~\cite{Mai:2014uma} as well as data on kaonic hydrogen~\cite{Bazzi:2011zj} 
one can fix the isovector part of the antikaon-nucleon amplitude at the threshold. 
Alternatively, a new measurement of the $K^{-}p \to \pi^{0}\Lambda$ reaction 
at as low energies as possible is highly desired as well.

\section{Dynamically generated poles}
\label{sec:poles}

Before proceeding further, we would like to make one more remark.
When the parameters of the models were fitted to experimental data the authors
used a coupled channels space composed of the physical meson-baryon states, 
10 channels in total that include all possible charge combinations that couple 
to the $K^{-}p$ state. This way the channel thresholds are treated properly 
in relation to the threshold effects observed in the $K^{-}p$ cross sections.
However, here we will look at pole movements and a large number of involved 
channels seriously complicates the matters especially when the poles move from one 
Riemann sheet (RS) to another one. Thus, to simplify our analysis quite a bit we will 
consider separately the meson-baryon states of different isospin. It means that 
we will use averaged hadron masses for the isospin multiplets instead of the physical 
masses of the charged particles that were used in the fits. We have checked that 
this change in treatment of masses has a minor impact on the energy dependence 
of the resulting transition amplitudes and on the reaction cross sections.

Let us now switch our attention to the pole content of the coupled channel chiral models 
introduced in the Section~\ref{sec:models}. It is well known that resonances observed in reaction 
cross sections and in the transition amplitudes can be related to the poles of the scattering 
S-matrix on unphysical RSs. A typical example is represented by the $\Lambda(1405)$ 
that is generated dynamically by the chiral models. There, it was found that two poles appear 
on the RS connected to the physical region at the real axis between the $\pi\Sigma$ 
and $\bar{K}N$ thresholds. In a notation we will adhere to in our paper this RS is denoted 
as $[-,+,+,+]$ with the signs marking the signs of the imaginary parts of the meson-baryon 
CMS momenta, see Eq.~\eqref{eq:G0}, in all four isoscalar coupled channels 
(unphysical for the $\pi \Sigma$ and physical for the $\bar{K}N$, $\eta \Lambda$ 
and $K\Xi$ ones). Similarly, in the isovector sector with five channels 
$\pi \Lambda$, $\pi \Sigma$, $\bar{K}N$, $\eta \Sigma$ and $K\Xi$ the RS connected 
with physical region in between the $\pi\Sigma$ and $\bar{K}N$ thresholds 
is denoted as $[-,-,+,+,+]$\footnote
{In the adopted notation the channels as well as the pertinent 
signs of the imaginary parts of the CMS momenta are ordered according to the channel 
threshold energies. The physical RSs in the isoscalar and isovector sectors 
are denoted as $[+,+,+,+]$ and $[+,+,+,+,+]$, respectively. By crossing the real axis
at any energy one gets to the RS with signs reverted for all channels that have 
thresholds below that energy.}. 

Many years ago it was already realized 
by Eden and Taylor \cite{Eden:1964zz} and later demonstrated by Pearce and Gibson \cite{Pearce:1988rk} 
that the origin of dynamically generated poles can be traced to the so called zero coupling 
limit (ZCL) in which the inter-channel couplings are switched off. In our work we utilize
their ideas and apply it to the system of channels coupled to the $\bar{K}N$ system. 
As mentioned in the Introduction,  the concept of switched off 
inter-channel couplings was also used by Hyodo and Weise \cite{Hyodo:2007jq} 
in their discussion of the two-pole structure of the $\Lambda(1405)$ resonance.

The basic idea employed in our analysis of dynamically generated resonances is as follows. 
The exact position of any pole is determined by the inter-channel couplings and 
the meson-baryon amplitudes obtained as solutions of the coupled channels equations 
are analytical with respect to those couplings. Any gradual change of the couplings 
leads to a continuous variation of the amplitudes and of the pole positions. If we gradually 
switch off the inter-channel couplings while keeping the diagonal couplings intact, 
the poles will move from their positions related to physically observed resonances 
to their positions in the ZCL. Thus, if there is a pole related to a resonance there 
must exist a related (connected by a continuous trajectory) pole in the ZCL and vice versa. 
To determine a position of the pole in ZCL related to the one found in the physical limit 
(with inter-channel couplings turned on) one can just follow the trajectory from 
the physical limit to the ZCL. Alternatively, one can start from any pole found 
in the ZCL and follow its trajectory while gradually turning on the non-diagonal inter-channel
couplings. Since the poles can evolve from the ZCL on various RSs one gets 
several trajectories of shadow poles, all having the same origin. As the inter-channel 
couplings are increasing (more generally, as any model parameters are varied) the evolving 
poles may move from one RS to another one by crossing the real axis. Some of the poles 
may end up at RSs very far from the physical one or very far from the real axis, 
so they hardly affect physical observables. On the other hand, it might happen 
that two (or even more) shadow poles are moved to positions where both of them affect 
physical observables and can be related to resonant states. An example of such a case 
was shown in Ref.~\cite{Cieply:2013sya} where both the $N^{\star}(1535)$ and the $N^{\star}(1650)$ 
resonances were generated dynamically by the model and found to originate from the same 
pole in the ZCL. A detailed analysis of pole movements upon varying the inter-channel 
couplings was also performed in Ref.~\cite{Pearce:1988rk} where the authors concentrated 
on the baryon-baryon interactions. 

It should be noted that a meaning of the {\it physically relevant pole} is rather fuzzy 
and arbitrary, especially when speaking about poles in the ZCL. In general, we looked 
at poles that are in the physical limit as far as about 250 MeV from the real axis and 
may appear in the ZCL even below the lowest $\pi\Lambda$ threshold. Sometimes the inter-channel 
couplings move the pole from its ZCL position to the physical one by quite a long trajectory, 
some other times we consider just the existence of the pole (though relatively far from a physical region) 
to be interesting as its exact position in the physical limit may not be restricted much 
by the fitted experimental data and it may become more relevant with a different parameter 
setting (at another possible $\chi^{2}$ minimum).

\subsection{Zero coupling limit}
\label{sec:ZCL}

In what follows, we wish to demonstrate how pole positions in the ZCL can be calculated. 
To simplify the discussion we start from the notation used for the Prague models.
The transition amplitudes matrix $f_{ij}$ has poles for complex energies $z$ 
(equal to the meson-baryon CMS energy $\sqrt{s}$ on the real axis) if a determinant 
of the inverse matrix is equal to zero, 
\begin{align}\label{eq:det}
{\rm det}|F^{-1}(z)| = {\rm det}|V^{-1}(z) - G(z)| = 0  \,,
\end{align}
where the potential matrix $V_{ij}$ is proportional to a coupling matrix $C_{ij}$ 
introduced in Eq.~(\ref{eq:vij}), 
and the intermediate state Green functions (meson-baryon loops) form the 
diagonal $G$ matrix. In the hypothetical ZCL, 
in which the non-diagonal inter-channel couplings are switched off ($V_{ij} = 0$ for $i \neq j$), 
the condition for a pole of the amplitude becomes
\begin{align}\label{eq:ZCL}
\prod_n [1/V_{nn}(z) - G_{n}(z)] = 0  \,,
\end{align}
where the index $n$ runs over all available coupled channels.
In such a scenario the RSs of different channels decouple and only two sheets exist per channel.
There will be a pole in channel $n$ at a RS [$+/-$] (physical/unphysical) 
if the pertinent $n$-th factor of the product on the l.h.s.~of Eq.~(\ref{eq:ZCL}) 
equals zero. For the separable Prague models the pole in the $n$-th channel exists 
if $z$ solves the equation
\begin{align}\label{eq:ZCLa}
\frac{4\pi f_{n}^{2}}{C_{nn}(z)}\: \frac{z}{M_n}
+ \frac{(\alpha_n + {\rm i}k_n)^{2}}{2\alpha_{n}}\:[g_{n}(k_{n})]^{2} = 0 \;\;\; ,
\end{align}
in which the sign of $\Im k_{n}$ relates to the appropriate RS. Alternatively, for the Murcia 
and Kyoto-Munich on-shell approaches the condition for a pole becomes
\begin{align}\label{eq:ZCLb}
\frac{1}{V_{nn}(z)} + \frac{1}{(4\pi)^{2}} \left[ a_{n}(\mu) 
+ 2\log \frac{M_{n}}{\mu} \right] + G_{0}(z,m,M) = 0  \;\;\; .
\end{align}
Thus, it is apparent that only states with nonzero diagonal couplings $C_{i,j=i} \neq 0$ can 
generate poles in the zero coupling limit. Anticipating that the WT term represents a major 
contribution to the meson-baryon interactions a quick glance at the structure of the SU(3) 
coefficients given in the first row of the Table \ref{tab:CNLO} implies non-zero matrix elements $V_{i,j=i}$ 
for the $\pi\Sigma$, $\bar{K}N$ and $K\Xi$ channels in both isospin sectors, $I=0$ and $I=1$. For each of these channels 
either of the Eqs.~(\ref{eq:ZCLa}) and (\ref{eq:ZCLb}) can provide us with one or more solutions, 
though some of them are not relevant for physics and appear to have only mathematical meaning. 
In principle, the contribution of the Born and NLO terms to the effective potential $V_{ij}$ 
leads to nonzero diagonal couplings in the remaining $\pi\Lambda$, $\eta\Lambda$ and $\eta\Sigma$ 
channels as well. In most cases, such contributions are small and the poles obtained as solutions 
of Eq.~(\ref{eq:ZCLa}) or (\ref{eq:ZCLb}) are too distant to affect physical observables when 
the inter-channel couplings are turned on. However, the Table \ref{tab:CNLO} demonstrates 
that the experimental data on their own do not exclude models with sizable NLO contributions. 
Then, the poles that emerge in the ZCL in these channels may be moved by inter-channel dynamics 
to positions related to the observed resonances.

It is instructive to look briefly at possible solutions of Eq.~(\ref{eq:ZCLa}) that can be sorted 
out in a clear manner. The equation can be transformed into a polynomial one in $z$ with 
a degree of 8, so it has exactly 8 solutions. It turns out that half of these solutions have $\Re z > 0$ and the other half have unphysical negative 
energies with $\Re z < 0$, so the later can be discarded. From the 4 remaining solutions two can 
be written as a sum of the baryon and meson energies, $z = E(M)+E(m)$, while the other two as $z = E(M)-E(m)$. 
Thus, Eq.~(\ref{eq:ZCLa}) has only two physically feasible solutions that comply 
with $z = E(M)+E(m)$ and $\Re z > 0$. Finally, since the Eq.~(\ref{eq:ZCLa}) is symmetric 
under the complex conjugation transition $z \to z^{*}$ the two solutions must both lie 
on the real axis or be represented by two conjugate solutions $z$ and $z^{*}$.

One may also ask a question what is a necessary condition to generate a ZCL bound state 
in a given channel,  i.e.~a pole at the real axis below the threshold energy $z_i({\rm th}) 
= M_i + m_i$ on the physical RS. Assuming that the couplings $C_{ii}$ and the meson decay constant $f_{i}$
are given, the Eq.~\eqref{eq:ZCLa} provides a condition for the inverse range $\alpha_{i}$.
It is easy to show that to form a bound state, the $\alpha_{i}$ must be larger than a minimal value to 
satisfy the condition
\begin{align}\label{eq:alphaM}
\alpha_{i} > \alpha_i({\rm min}) = \frac{16\pi}{\tilde{C}_{ii}}\,\frac{f_{i}^{2}}{\omega_{i}}\,,
\end{align}
where $\omega_{i}$ stands for a meson-baryon reduced mass and the chiral couplings 
$\tilde{C}_{ii}$ are the threshold values given in the Table \ref{tab:CNLO}. 
Considering only the WT interaction in the $\bar{K}N(I=0)$ 
channel, the condition is $\alpha_{\bar{K}N} \gsim 600$ MeV. Thus, the parameters $\alpha_i$ fitted 
to the experimental data must satisfy this condition to provide a bound state in the ZCL to be 
turned into a dynamically generated resonance state when the inter-channel couplings are switched on.

It is more difficult to make any predictions for solutions of the Eq.~(\ref{eq:ZCLb}) that involves 
the meson-baryon loop regularized by dimensional regularization. As the equation cannot be transformed 
into a polynomial one, a number of its solutions is not fixed and depends on a specific interplay 
of the model parameters in a given channel. For positive complex energies $\Re z > 0$ the equation 
may generate either two complex conjugate solutions above the channel threshold ($\Re z > M_i+m_i$) 
or one to two solutions at the real axis below the threshold. Since the later may appear at 
unphysical energies  $\Re z < M_i$, the Eq.~(\ref{eq:ZCLb}) may have no physically relevant 
solutions. We will demonstrate  the situation for a specific case discussed in the following section.

\subsection{Weinberg-Tomozawa models}
\label{sec:WTmod}

It is easiest to start our discussion of the pole content generated by various models 
by looking at the simplest case when the interaction kernel is reduced to the WT term. 
Although the NLO approaches are superior to the WT ones in terms of data reproduction,
they introduce a large number of additional interaction terms and related parameters 
fitted to the data which may blur the analysis we aim at here. As far as 
the WT interaction dominates the low energy $\bar{K}N$ interactions the results achieved with 
such models should give us a good guidance when discussing the more complex models. It is also natural 
to perform our analysis in the basis of meson-baryon states with a particular isospin, separately 
for the isoscalar ($I=0$) and isovector ($I=1$) states derived from the physical channels by applying 
standard isospin projections. This reduces vastly the number of RSs considered in the analysis 
and allows for a straightforward assignment of any physically relevant poles to resonances 
of a particular isospin.

Among the theoretical models adjusted to reproduce the most recent experimental data on 
kaonic hydrogen,  the 1s level shift and width measured by the SIDDHARTA collaboration 
\cite{Bazzi:2011zj}, there are only two  papers \cite{Cieply:2011nq}, \cite{Ikeda:2012au} 
presenting results with meson-baryon interactions restricted 
to the WT term. The pertinent models are the \CSWT{} model \cite{Cieply:2011nq} and 
the \IHWWT{} model \cite{Ikeda:2012au}  introduced in Section~\ref{sec:models}. 
The physically relevant solutions of either Eq.~(\ref{eq:ZCLa}) 
for the \CSWT{} model or Eq.~(\ref{eq:ZCLb}) for the \IHWWT{} model are listed in 
Table~\ref{tab:ZCL-WT}.

\begin{table}[h]
\caption{The pole positions in the zero coupling limit are presented in a form that shows 
their complex energy $z$ in round brackets and the Riemann sheet the pole is found on 
in the square brackets. The characters of the states corresponding to the poles are specified 
as well as resonant (R), bound (B) and virtual (V) ones.}
\begin{center}
\begin{tabular}{|cc|lc|lc|}
\hline
         &       & \multicolumn{2}{c|}{\CSWT{} model} & \multicolumn{2}{c|}{\IHWWT{} model} \\
\hline
 isospin & channel   & $z$(MeV)[$+/-$] &  status & $z$(MeV)[$+/-$] &  status \\ 
\hline
   & $\pi\Sigma$ &  (1366,  $-$91)$[-]$ & R  &  (1364, $-$100)$[-]$ & R   \\
 0 & $\bar{K}N $ &  (1434,    0)$[+]$ & B  &  (1419,    0)$[+]$ & B   \\
   & $K\Xi$      &  (1809,    0)$[+]$ & B  & \multicolumn{1}{c}{---} & - \\
\hline
   & $\pi\Sigma$ &  (1387, $-$220)$[-]$ & R  &  (1423, $-$196)$[-]$ & R   \\
 1 & $\bar{K}N $ &  (1158,  $-$18)$[-]$ & R  & \multicolumn{1}{c}{---} & - \\
   & $K\Xi$      &  (1658,    0)$[-]$ & V  & \multicolumn{1}{c}{---} & - \\ 
\hline
\end{tabular}
\end{center}
\label{tab:ZCL-WT}
\end{table}

\begin{figure}[thb]
\centering
\resizebox{0.6\textwidth}{!}
{
\includegraphics{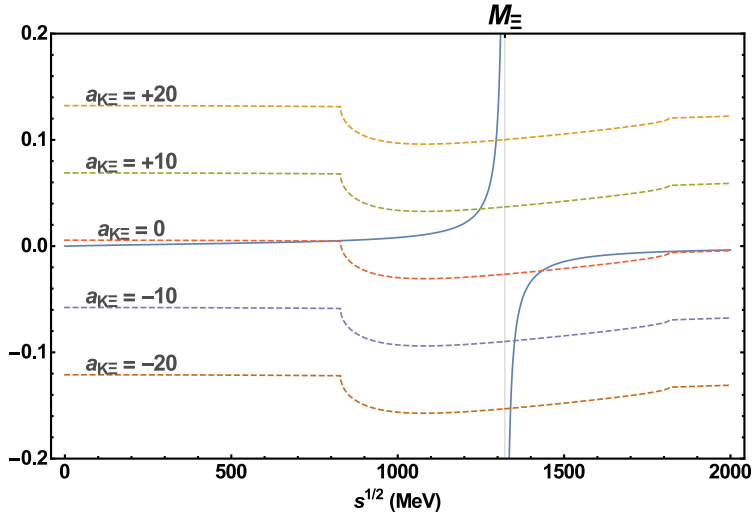}
}
\caption{Solutions of the ZCL equation Eq.~(\ref{eq:ZCLb}) are found as crossings of 
the $1/V$ plot  and the meson-baryon loop function $G$ calculated for several values 
of the subtraction constant $a_{K\Xi}$ on the unphysical RS. 
The thin vertical line shows the $1/V$ asymptote at the energy $\sqrt{s} = M_{\Xi}$.}
\label{fig:ZCLb}
\end{figure}

As we see the \CSWT{} model provides us with a solution of the ZCL equation for all three meson-baryon 
channels that have non-zero $C_{ii}$ coupling and in both the $I=0$ and $I=1$ sectors. On the 
other hand, the pole content of the \IHWWT{} model is limited only to the already well known 
isoscalar $\pi\Sigma$ 
and $\bar{K}N$ poles \cite{Hyodo:2007jq} and to the isovector $\pi\Sigma$ resonance that is very far 
from the real axis. The fact that the \IHWWT{} model does not generate any additional poles that 
appear in the \CSWT{} model is related to the values of the subtraction constants fitted 
to experimental data in \cite{Ikeda:2012au} as demonstrated in Figure \ref{fig:ZCLb}. There, 
in the isoscalar $K\Xi$ channel $1/V_{K\Xi}(\sqrt{s})$ part of Eq.~(\ref{eq:ZCLb}) is plotted 
against the one-loop integral $G_{K\Xi}(\sqrt{s})$ on the unphysical RS and for several values 
of the subtraction constant $a_{K\Xi}$. The intersection of both curves gives back the energy at which 
a pole occurs in the ZCL. Note also that $1/V$ asymptote at the energy $\sqrt{s} = M_{\Xi}$ separates 
the unphysical region of the CMS energies $\sqrt{s} < M_{\Xi}$ from the one in which physically
relevant poles can be found. It is apparent that the subtraction constant must be smaller than 
a certain value to generate a crossing of the $1/V$ and $G$ lines in the $\sqrt{s} > M_{i}$ region. 
For the discussed case the \IHWWT{} value\footnote
{The value $0.039$ given in Ref.~\cite{Ikeda:2012au} should be multiplied by a factor $16\pi^{2}$ 
to comply with our definition of the subtraction constant in Eq.~(\ref{eq:MBloop}).}
$a_{K\Xi} = 6.23$ reported in Ref.~\cite{Ikeda:2012au}
does not generate a physically relevant solution of the Eq.~(\ref{eq:ZCLb}). 
We have checked numerically that the solution found at $\sqrt{s} < M_{\Xi}$ 
moves to even lower energies when the inter-channel couplings are switched on. This finding is 
relevant for any other dynamically generated resonances provided by the {\it on-shell models}. 
Since the poles in the physical limit may exist only if a related pole is found in the ZCL, 
the subtraction constant in the pertinent channel (where the pole originates) must be smaller 
than a certain maximal value $a_i < a_i({\rm max})$. This represents a similar condition 
as the one given by Eq.~\eqref{eq:alphaM} for the separable potential model. For exactly 
the same reason as there, if the condition is not met, the pole may not originate 
from such meson-baryon channel. Similarly, because the fitted values of the subtraction 
constants are larger than the maximal ones, the \IHWWT{} model does not generate 
poles in the ZCL for the isovector channels $\bar{K}N$ and $K\Xi$ neither. Since the experimental 
data used in the fits apparently do not put serious restrictions on these channels it is plausible 
to accept that different models come to differing conclusions concerning the existence of poles 
generated by the diagonal couplings in those channels. 

Once we find the pole positions in the ZCL we can follow their movements on the complex energy 
manifold by gradually turning on the inter-channel couplings. We have done it by scaling 
the non-diagonal couplings by a factor $x$ that ranges from $0$ in the ZCL to $1$ 
in the physical limit. Since the scattering matrix is analytical in $x$, no pole may disappear 
when evolving from its position found for $x=0$. In the multiple channel setup each pole found 
in the ZCL  generates a multiplet of shadow poles evolving from the same original position at 
various RSs. 
The pole that is nearest to the physical region plays a dominant role in terms of having an impact 
on physical observables though it may happen that two (or even more) shadow poles are about equally 
away from the physical RS and there is no way to say which of them is dominant. 
We also note that physical observables may be dominated by poles at a more distant RSs 
than those reached by crossing the real axis from the physical region\footnote{
The RS reached by crossing the real axis in between two thresholds is often called {\it the second RS},
though it is a bit misleading when more than two channels and multiple thresholds are involved and 
there is more than one such RS.
}.
Typically, such poles exist close to a threshold on a RS reached from the physical region by turning 
around the threshold. What matters is the {\it length} of a shortest path from the pole to a given 
point at the real axis in the physical region at which the physical observables are evaluated. 
Sometimes a pole at the {\it second RS} is more distant and affects physics less than another 
pole at more distant RS, so it is important to look for poles at various RSs, not only on those
reached by crossing the real axis from the physical region.

\begin{figure}[thb]
\resizebox{0.5\textwidth}{!}{%
  \includegraphics{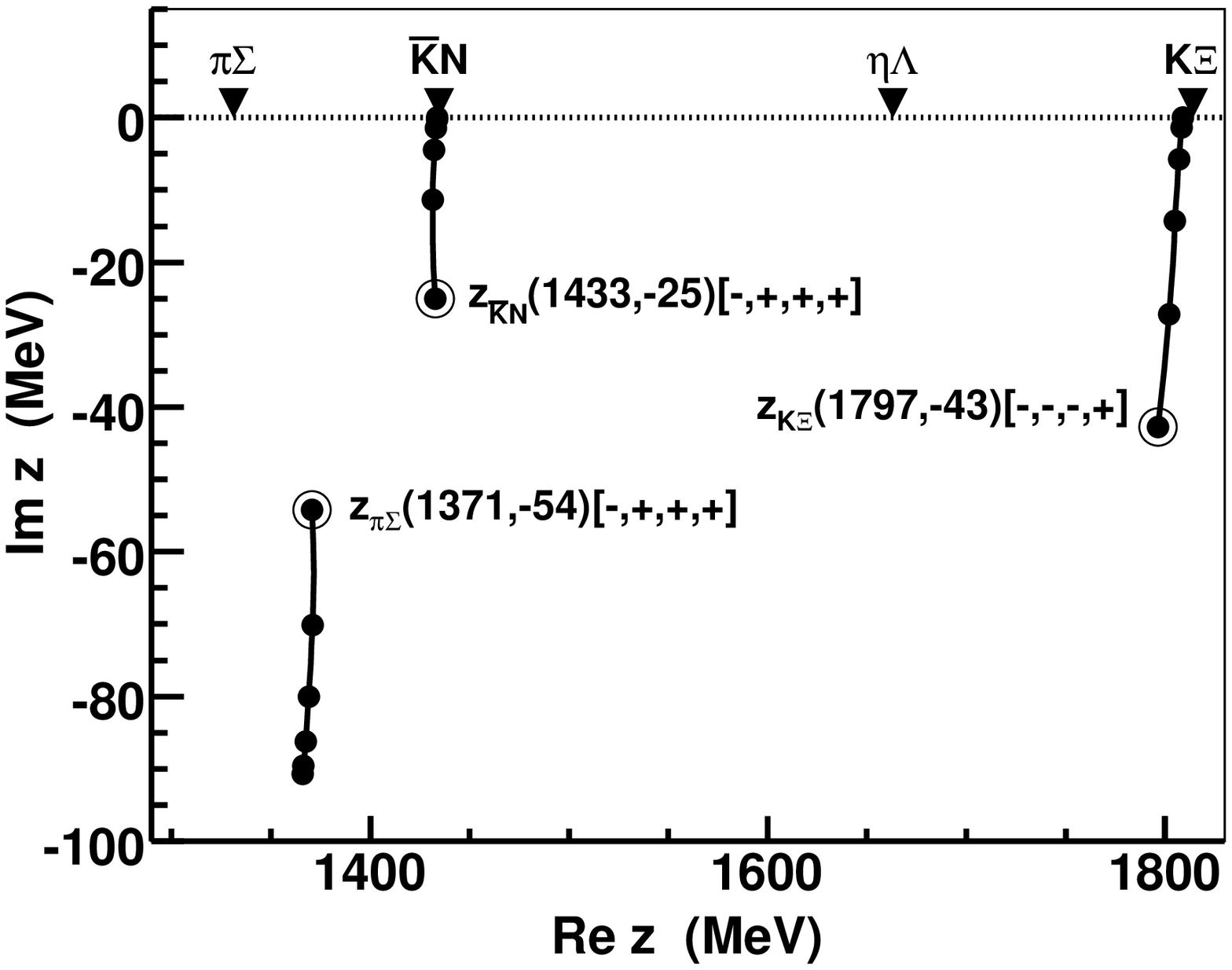}
}
\resizebox{0.5\textwidth}{!}{%
  \includegraphics{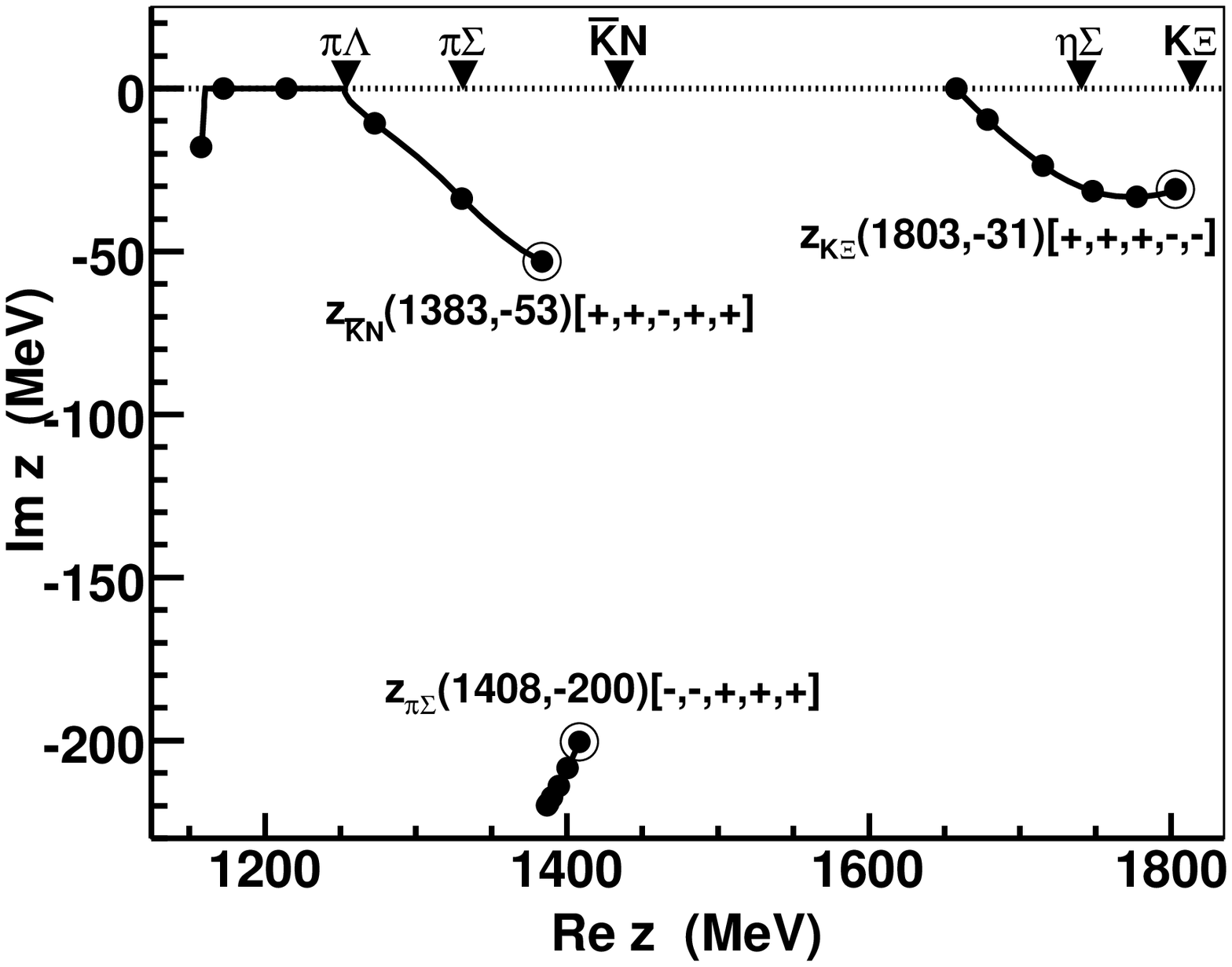}
}\\
\resizebox{0.5\textwidth}{!}{%
  \includegraphics{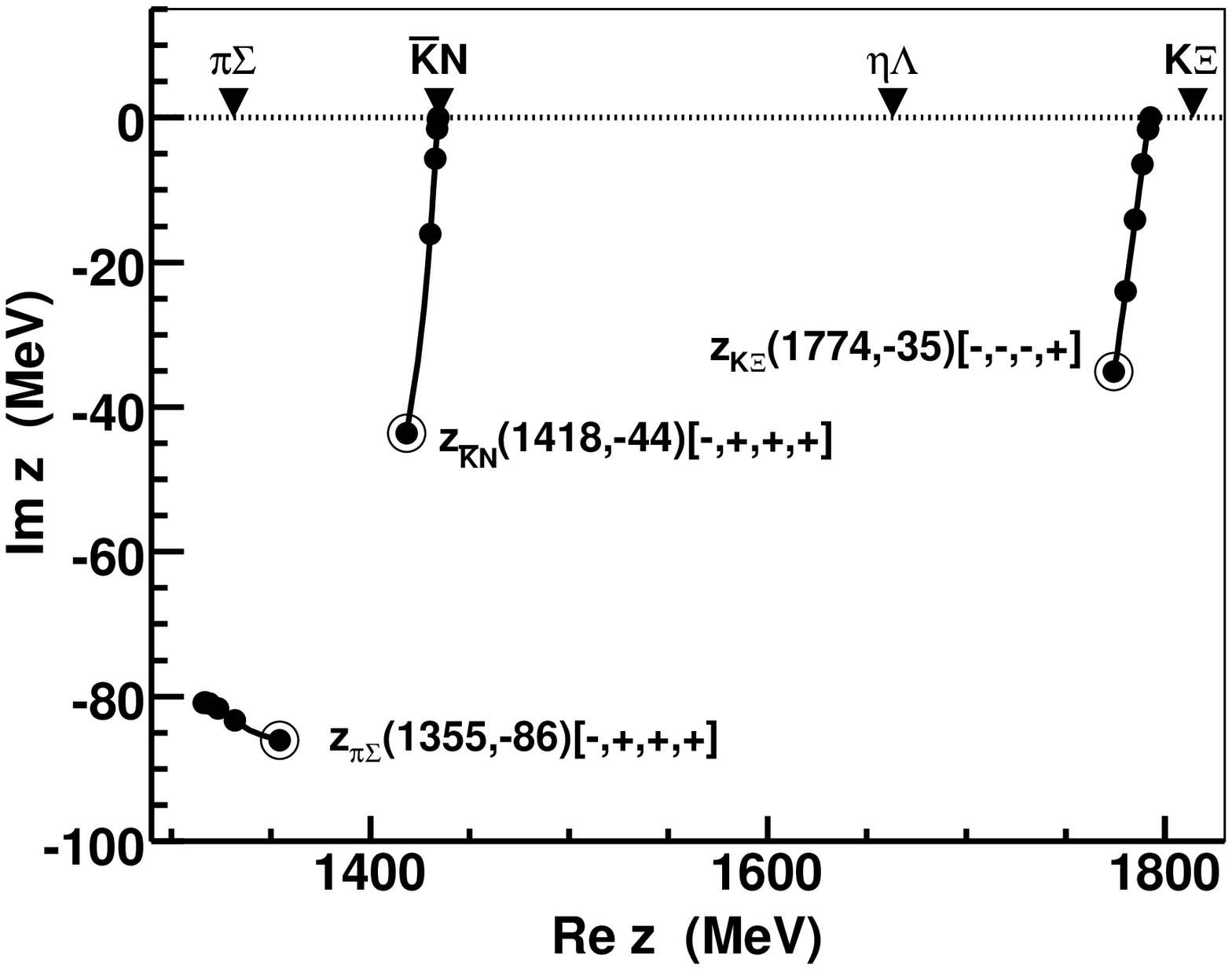}
}
\resizebox{0.5\textwidth}{!}{%
  \includegraphics{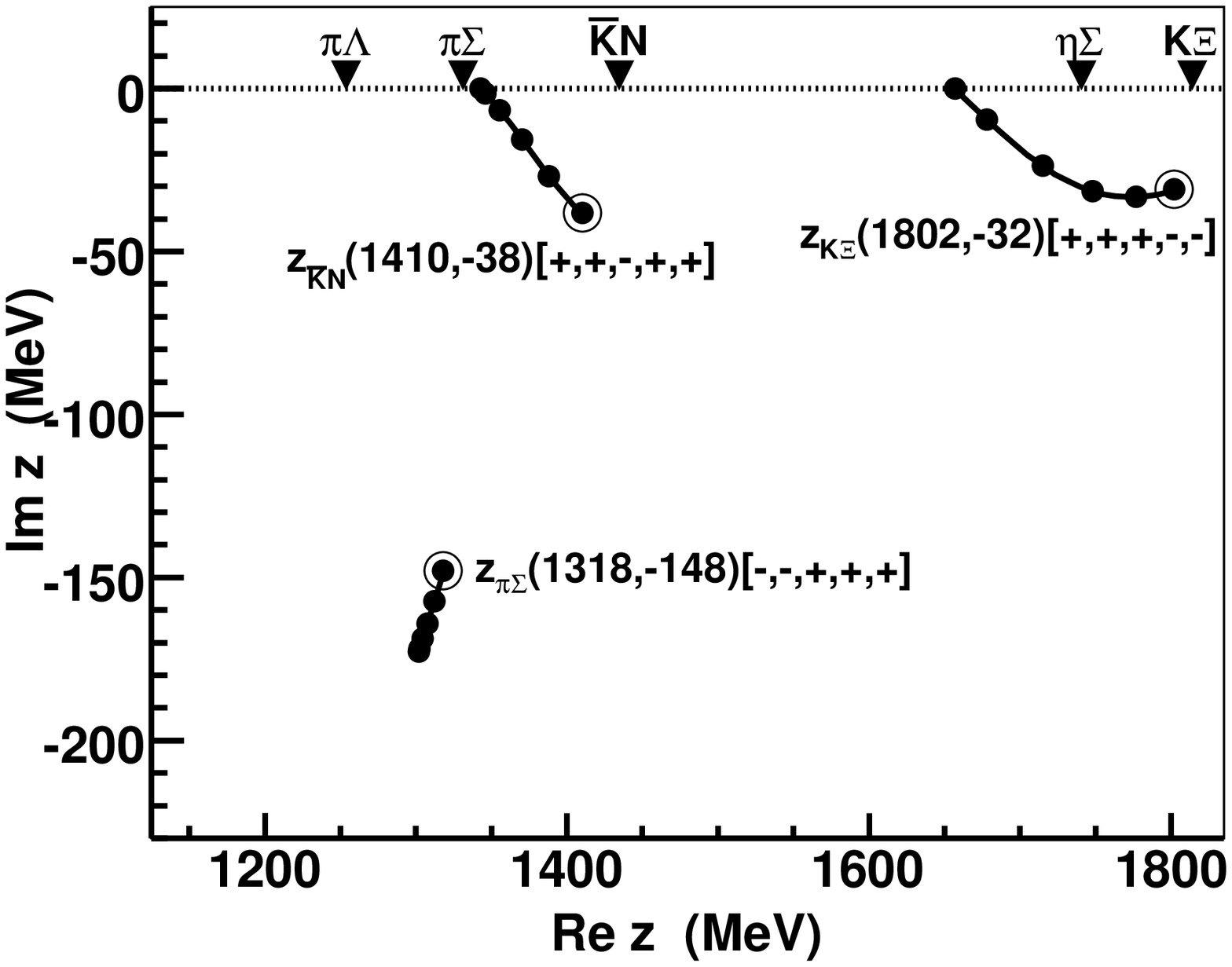}
}
\caption{The \CSWT{} (top panels) and \CSnlo{} (bottom panels) models. Trajectories of the poles 
related to the $\pi\Sigma$, $\bar{K}N$ and $K\Xi$ channels are obtained by scaling the non-diagonal 
inter-channel couplings. The left (right) panel relates to the isoscalar (isovector) channels. 
The dots on the trajectories indicate pole positions for the scaling factor from $x=0$ 
to $x=1$ in steps of $0.2$ with the pole positions in the physical limit emphasized by large empty circles.
The triangles at the top of the real axis indicate the channel thresholds.
}
\label{fig:ZCL-CS}       
\end{figure}

The trajectories 
of the poles that are dominant in the physical limit are visualized in Figure~\ref{fig:ZCL-CS}
that shows the results obtained with the Prague models. We have assigned the observed poles to 
the channels in which they persist in the ZCL and also specify the RS on which the pole evolves 
and its final complex energy position in the physical limit. For the \CSWT{} model, the ZCL pole 
positions are those as given in Table~\ref{tab:ZCL-WT}.

First, we look at the top panels of Figure~\ref{fig:ZCL-CS} and discuss the pole trajectories obtained 
with the \CSWT{} model. The isoscalar poles that evolve from the $\pi\Sigma$ resonance and from 
the $\bar{K}N$ bound state for $x=0$ are the two poles that are usually related to the 
$\Lambda(1405)$ resonance. As anticipated, in the physical limit (for $x=1$) both poles appear 
on the RS that can be reached from the physical region by crossing the real axis in-between 
the $\pi\Sigma$ and $\bar{K}N$ thresholds. The third isoscalar 
pole $z_{K\Xi}$ can be related to the $\Lambda(1670)$ resonance, though its position is quite off 
the one observed in experimental spectra. We should stress that the \CSWT{} model is 
not expected to work 
so well at energies hundreds MeV away from the $\bar{K}N$ threshold, so the discrepancy is of little 
concern here. The poles found in the isovector sector are not so well known, only the $\bar{K}N$ one 
was observed earlier, too, see Refs.~\cite{Oller:2000fj, Jido:2003cb}. It is understood that it 
relates to the cusp structure in the energy dependence of the elastic $K^{-}n$ amplitude obtained 
for both, the Prague and the Kyoto-Munich models. The $\pi\Sigma$ isovector pole is too far from the real axis, 
so it cannot affect physical observables. The $K\Xi$ isovector pole may be related to the 
$\Sigma(1750)$, a three star resonance.

Although we do not present a similar figure with pole trajectories for the Kyoto-Munich models, we 
have checked that the three ZCL poles obtained for the \IHWWT{} model and listed in 
Table~\ref{tab:ZCL-WT}  evolve in a completely similar manner as those of the \CSWT{} model 
when the inter-channel couplings are switched on. Particularly, the $\pi\Sigma$ and the 
$\bar{K}N$ isoscalar poles evolve into the two poles 
assigned to the $\Lambda(1405)$ reaching in the physical limit the pole positions reported 
in Ref.~\cite{Ikeda:2012au}. 
The isovector $\pi\Sigma$ pole found in the ZCL ends up (for $x=1$) at a pole position 
$z_{\pi\Sigma} = (1483, -126)$ MeV at the same $[-,-,+,+,+]$ RS where the pole is found 
for the \CSWT{} model. Thus, the exact positions of the three poles generated by both the 
\CSWT{} and \IHWWT{} 
models vary with the specific model for $x=0$ as well as for $x=1$, though the picture is 
not altered qualitatively. 
The fact that the Kyoto-Munich model does not generate poles in the other three channels can be explained 
by noting that the experimental data do not restrict much the subtraction constants related to those 
channels, so their values could easily be smaller (negative) without affecting much the general fit 
while generating more poles in the ZCL.

For a completeness we also mention that the isotensor ($I=2$) WT coupling for 
the $\pi\Sigma$ channel is also not equal to zero, $C_{\pi\Sigma, \pi\Sigma}^{I=2} = -2$. 
However, the ZCL solutions of Eqs.~(\ref{eq:ZCLa}) and (\ref{eq:ZCLb}) found 
for the considered models are too far from the real axis and thus do not affect 
physical observables.

\subsection{NLO models}
\label{sec:NLOmod}

Let us discuss how the inclusion of the NLO terms in the chirally motivated approaches changes 
the picture outlined in the previous section. First, we look at the NLO versions of the Prague 
and Kyoto-Munich approaches, Refs.~\cite{Cieply:2011nq} and \cite{Ikeda:2012au}. For both models, 
the pole positions in the ZCL are given in the Table~\ref{tab:ZCL-NLO}. Nothing is
really new in case of the \CSnlo{} model besides the fact that the positions of 
the poles were modified to some extent due to introduction of the NLO terms. 
The pole positions in the physical limit do not offer any surprises neither as we demonstrate 
in the bottom panels of Figure \ref{fig:ZCL-CS}. There, the only notable changes with respect 
to the observations made for the \CSWT{} model are shifts of the two $\Lambda(1405)$ 
poles further from the real axis (for $x=1$) and a more natural position of the isovector 
$\bar{K}N$ pole in the ZCL (for $x=0$).

\begin{table}[h]
\caption{The pole positions in the zero coupling limit for the \CSnlo{} and \IHWnlo{} models. 
See the caption of Table \ref{tab:ZCL-WT} and the text for more details.}
\begin{center}
\begin{tabular}{|cc|lc|lc|}
\hline
         &       & \multicolumn{2}{c|}{\CSnlo{} model} & \multicolumn{2}{c|}{\IHWnlo{} model} \\
\hline
 isospin & channel   & $z$(MeV)[$+/-$] &  status & $z$(MeV)[$+/-$] &  status \\ 
\hline
   & $\pi\Sigma$ &  (1316,  $-$81)$[-]$ & R  &  (1358, $-$114)$[-]$ & R   \\
 0 & $\bar{K}N $ &  (1434,    0)$[+]$ & B  &  (1411,    0)$[+]$ & B   \\
   & $K\Xi$      &  (1793,    0)$[+]$ & B  & \multicolumn{1}{c}{---} & - \\
\hline
   & $\pi\Sigma$ &  (1302, $-$173)$[-]$ & R  &  (1360, $-$260)$[-]$ & R   \\
 1 & $\bar{K}N $ &  (1343,    0)$[-]$ & V  & \multicolumn{1}{c}{---} & -  \\
   & $\eta\Sigma$ & \multicolumn{1}{c}{---} & - & (1420, 0)$[+]$ & B  \\
   & $K\Xi$      &  (1657,    0)$[-]$ & V  & \multicolumn{1}{c}{---} & -  \\
\hline
\end{tabular}
\end{center}
\label{tab:ZCL-NLO}
\end{table}

Concerning the \IHWnlo{} model, a new pole emerges from the $\eta\Sigma$ 
bound state found in the ZCL, though its exact position on the real axis 
is partly obscured by the left-hand cut of the Born u-term there. 
When the inter-channel couplings are switched on it develops into a pole 
in the physical limit located on the $[-,-,-,+,+]$ RS at the energy 
$z =$ (1420, -11) MeV, very close to the $\bar{K}N$ threshold. Most likely, 
the pole is responsible for the spike structure observed in the $K^{-}n$ 
amplitude in Figure \ref{fig:KNampl} playing the same role there 
as the $\bar{K}N(I=1)$ pole does for the \CSnlo{} model. The isoscalar 
sector of the \IHWnlo{} model offers no surprises with the pole content 
and dynamics of the model not changed when the NLO terms are introduced.
There, the variations of the pole positions when going from the \IHWWT{} 
to the \IHWnlo{} ones are even smaller than those for the Prague models.

\begin{figure}[thb]
\resizebox{0.5\textwidth}{!}{%
  \includegraphics{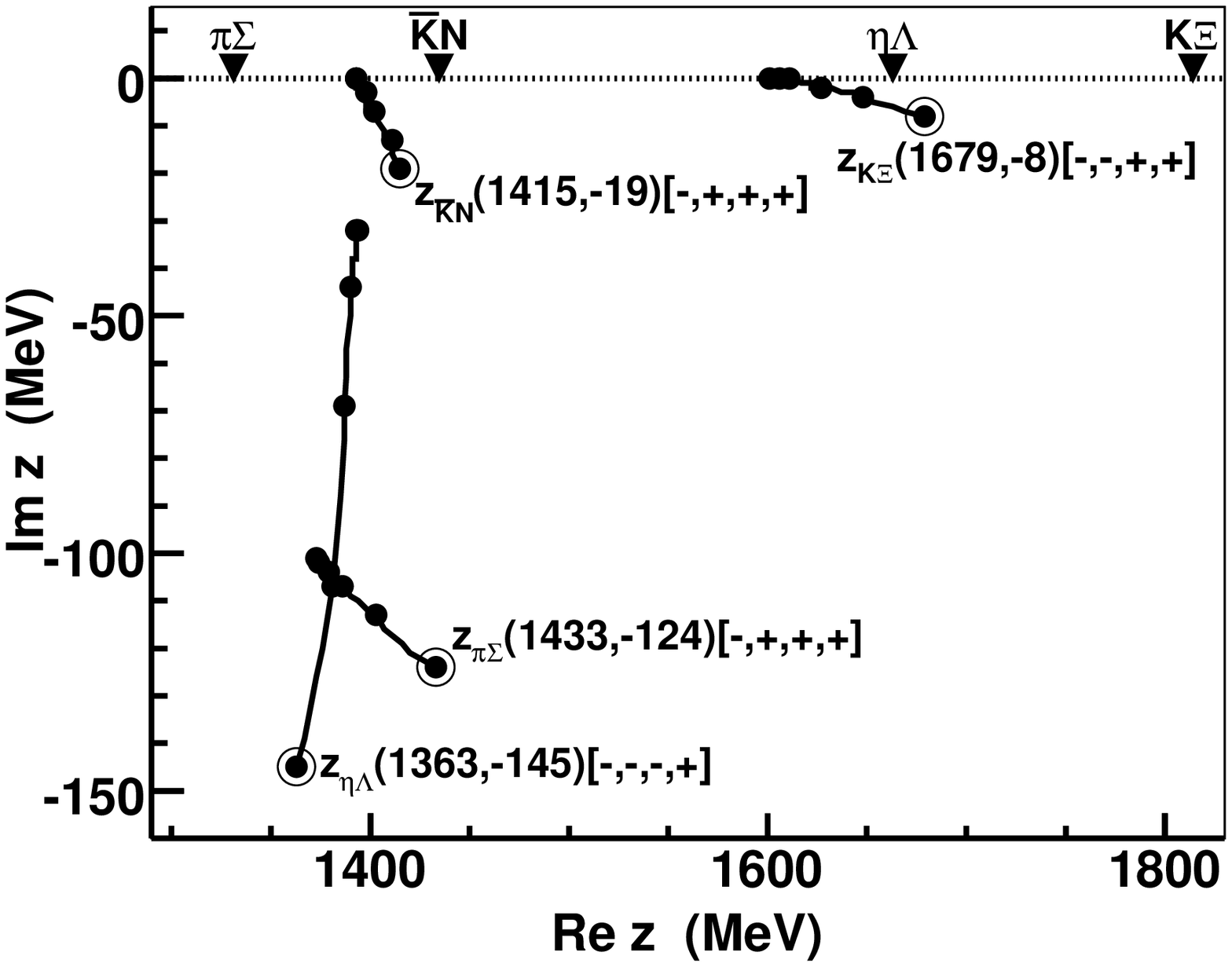}
}
\resizebox{0.5\textwidth}{!}{
  \includegraphics{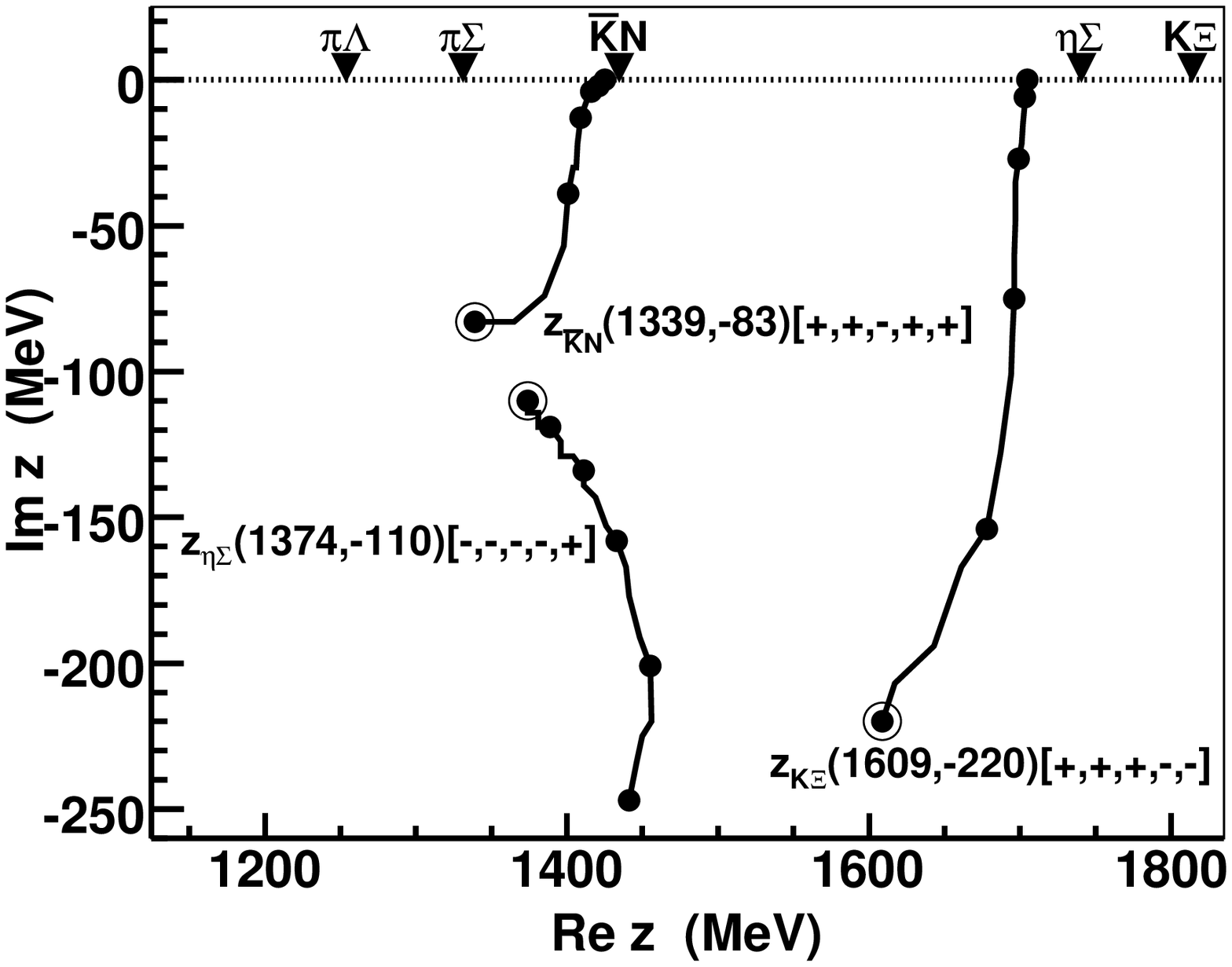}
}\\
\resizebox{0.5\textwidth}{!}{%
  \includegraphics{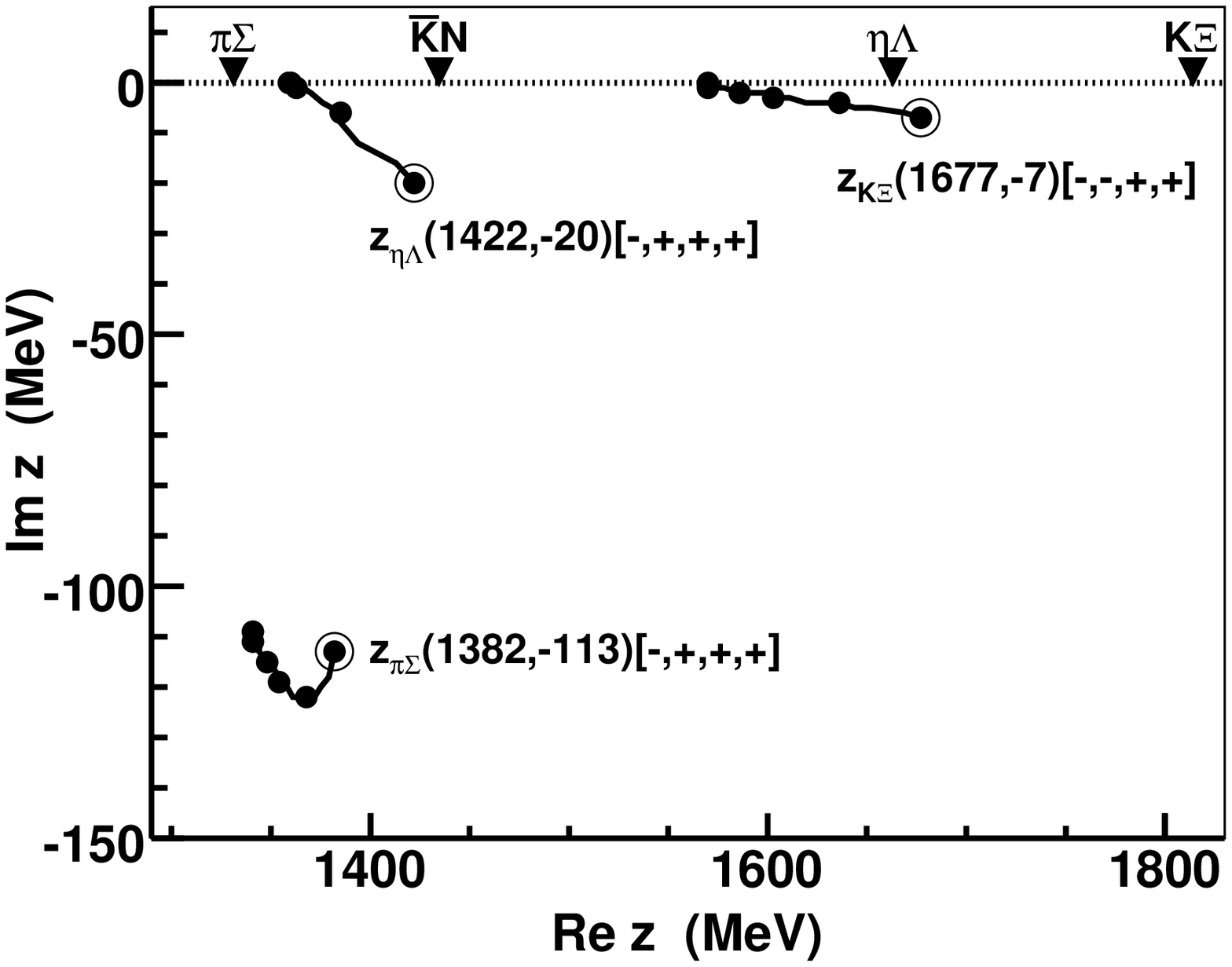}
}
\resizebox{0.5\textwidth}{!}{%
  \includegraphics{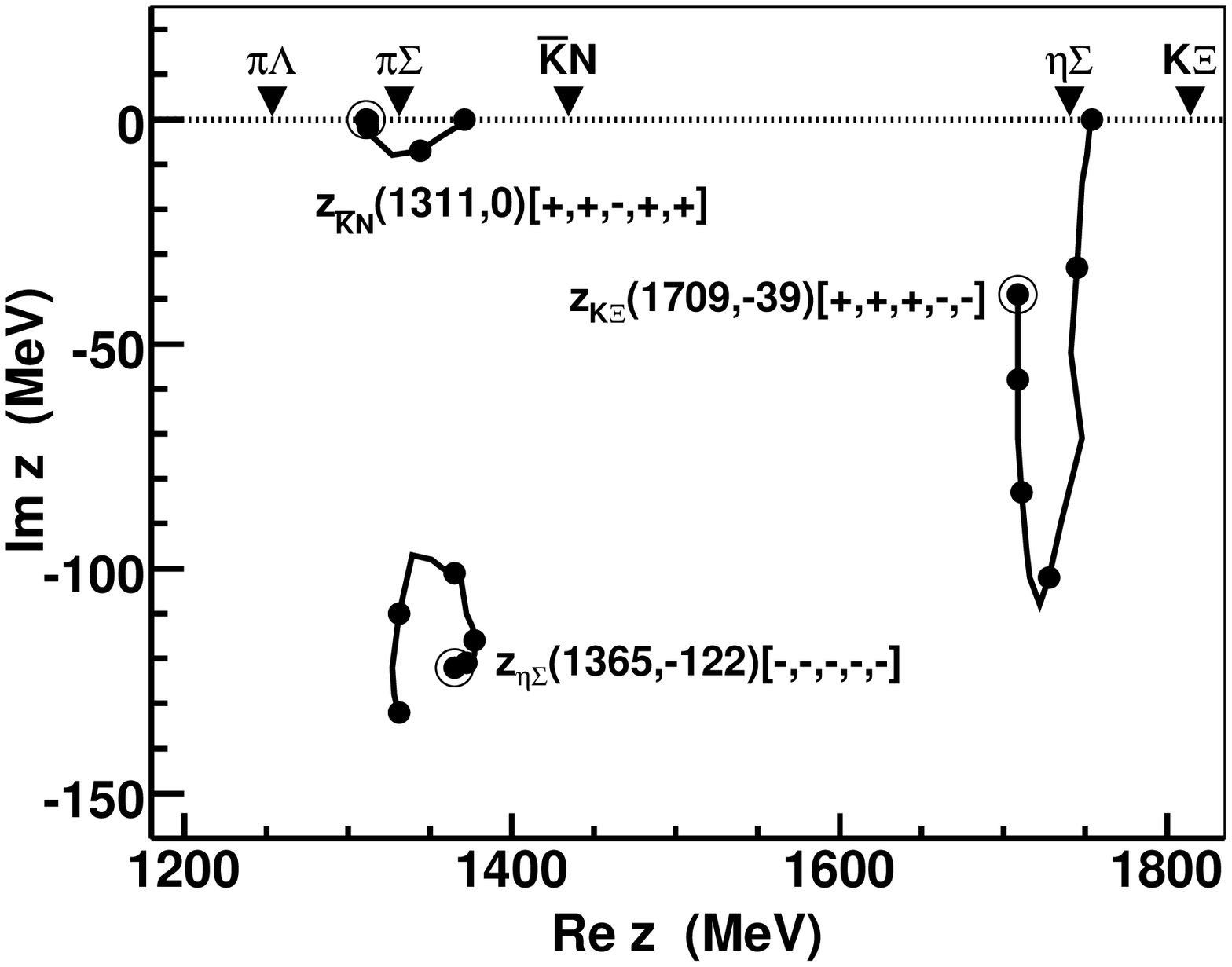}
}
\caption{Trajectories of the poles generated for the \GOi{} (top panels) and \GOii{} 
(bottom panels) models. The left (right) panel relates to the isoscalar (isovector) 
channels. For further notation, see Figure \ref{fig:ZCL-CS}.
}
\label{fig:ZCL-GO}       
\end{figure}

The Murcia models \cite{Guo:2012vv} offer a much richer pole content than the Prague and Kyoto-Munich models. 
This feature is caused by quite large NLO LECs that lead to sufficiently strong 
diagonal couplings even for those channels that could not generate any ZCL poles 
because the pertinent WT coupling was zero. Thus, in the ZCL we were able to locate 
solutions for channels including the $\pi\Lambda$, $\eta\Lambda$ and $\eta\Sigma$ ones, 
though some of them develop into positions that are rather far from the physical region. 
The ZCL positions of the poles that we found most relevant (or most interesting) 
are collected in Table~\ref{tab:ZCL-GO} and their trajectories are depicted in 
Figure~\ref{fig:ZCL-GO}. 
The most interesting point here is that for the \GOii{} model the higher $\Lambda(1405)$ 
pole originates from the $\eta\Lambda$ bound state and not from the $\bar{K}N$ one 
as anticipated and provided by the Prague, Kyoto-Munich and the \GOi{} models. The $\bar{K}N$ related 
pole is still there, though the \GOii{} model has it bound (in the ZCL) at much lower 
energy, below the $\pi\Sigma$ threshold. This large binding is apparently caused by 
a large coupling in the $\bar{K}N(I=0)$ channel, almost three times larger than provided 
by only the WT interaction. For the \GOi{} model this coupling is even larger, though 
it is compensated by a larger subtraction constant to generate the ZCL solution 
of Eq.~(\ref{eq:ZCLb}) at a higher energy. The different sets of parameters used 
in the \GOi{} and \GOii{} then lead to a different pole origins and different pole 
dynamics despite the fact that in the physical limit both models generate the higher 
$\Lambda(1405)$ pole at about the same position.

\begin{table}[h]
\caption{The pole positions in the zero coupling limit for the \GOi{} and \GOii{} models.
See the caption of Table~\ref{tab:ZCL-WT} and the text for more details.}
\begin{center}
\begin{tabular}{|cc|lc|lc|}
\hline
         &       & \multicolumn{2}{c|}{\GOi{} model} & \multicolumn{2}{c|}{\GOii{} model} \\
\hline
 isospin & channel   & $z$(MeV)[$+/-$] &  status & $z$(MeV)[$+/-$] &  status \\ 
\hline
   & $\pi\Sigma$   &  (1373, $-$101)$[-]$ & R  &  (1341, $-$109)$[-]$ & R   \\
 0 & $\bar{K}N $   &  (1393,    0)$[+]$ & B  &  (1289,    0)$[+]$ & B   \\
   & $\eta\Lambda$ &  (1394,  $-$32)$[-]$ & R  &  (1359,    0)$[+]$ & B   \\
   & $K\Xi$        &  (1601,    0)$[+]$ & B  &  (1570,    0)$[+]$ & B   \\
\hline
   & $\bar{K}N $   &  (1425,    0)$[-]$ & V  &  (1371,    0)$[-]$ & V   \\
 1 & $\eta\Sigma$  &  (1441, $-$247)$[-]$ & R  &  (1331, $-$132)$[-]$ & R   \\
   & $K\Xi$        &  (1705,    0)$[-]$ & V  &  (1754,    0)$[-]$ & V   \\ 
\hline
\end{tabular}
\end{center}
\label{tab:ZCL-GO}
\end{table}

The isoscalar $K\Xi$ related poles obtained by both Murcia models match nicely the position 
of the $\Lambda(1670)$ resonance. This achievement is due to the  use of additional 
experimental data at higher energies in the fits performed in Ref.~\cite{Guo:2012vv}, 
particularly the $K^{-}p \to \eta\Lambda$ reaction data. The Murcia models 
also generate the isovector $\bar{K}N$ and $K\Xi$ poles in correspondence with those 
generated by the Prague models. Additionally, the $\eta\Sigma(I=1)$ pole is worth noting, 
though it appears at a position rather distant from a physical region because 
it is located on the $[-,-,-,-,-]$ RS and not on the nearer $[-,-,+,+,+]$ one.

As the possibility of forming the higher $\Lambda(1405)$ pole from the $\eta\Lambda$ 
ZCL bound state  has not been anticipated in the literature, we feel that this requires 
an additional comment. The models restricted to the WT interaction do not 
have this option and as far as the NLO LECs are relatively small the situation is qualitatively 
not changed as manifested by the Prague and Kyoto-Munich NLO models. However, the experimental data used 
in the fits do not discriminate between models with small NLO contributions and those that 
have them comparable with the LO ones. The fact that the meson-baryon interactions in the 
$S=-1$ sector are quite well described by models employing only the WT term is a mystery 
which is not understood, so there is no reason to believe that the NLO LECs should be small. 
On the other hand, it is definitely more natural to assume that the $\Lambda(1405)$ is formed 
from the $\bar{K}N$ bound state that is transformed into a resonance (quasi-bound state) 
by acquiring a non-zero width due to couplings to other channels, most notably to the $\pi\Sigma$ 
one. For this reason, we consider the \GOi{} model a preferred option over the \GOii{} one.

Finally, we present our results obtained for the pole content and pole movements generated 
by the Bonn models of Ref.~\cite{Mai:2014xna}. There, we found it very difficult to follow 
properly the movement of some poles (especially in the isovector sector) since the models 
are not based on a potential concept and the scattering amplitudes are derived 
in a different manner as outlined in Section~\ref{sec:models}. The relevant 
ZCL poles obtained for the Bonn models are collected in Table~\ref{tab:ZCL-MM}.

\begin{figure}[t]
\resizebox{0.5\textwidth}{!}{%
  \includegraphics{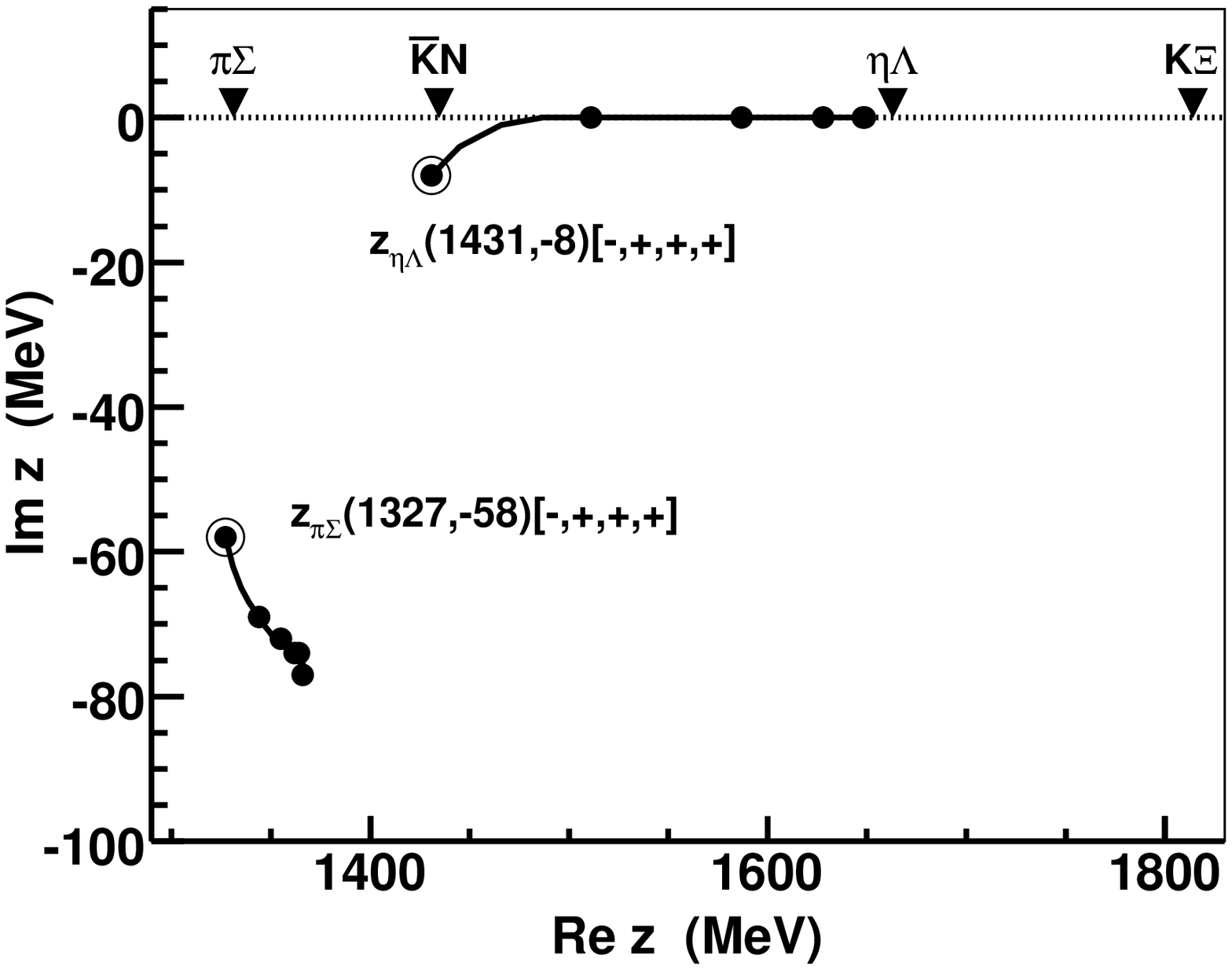}
}
\resizebox{0.5\textwidth}{!}{%
  \includegraphics{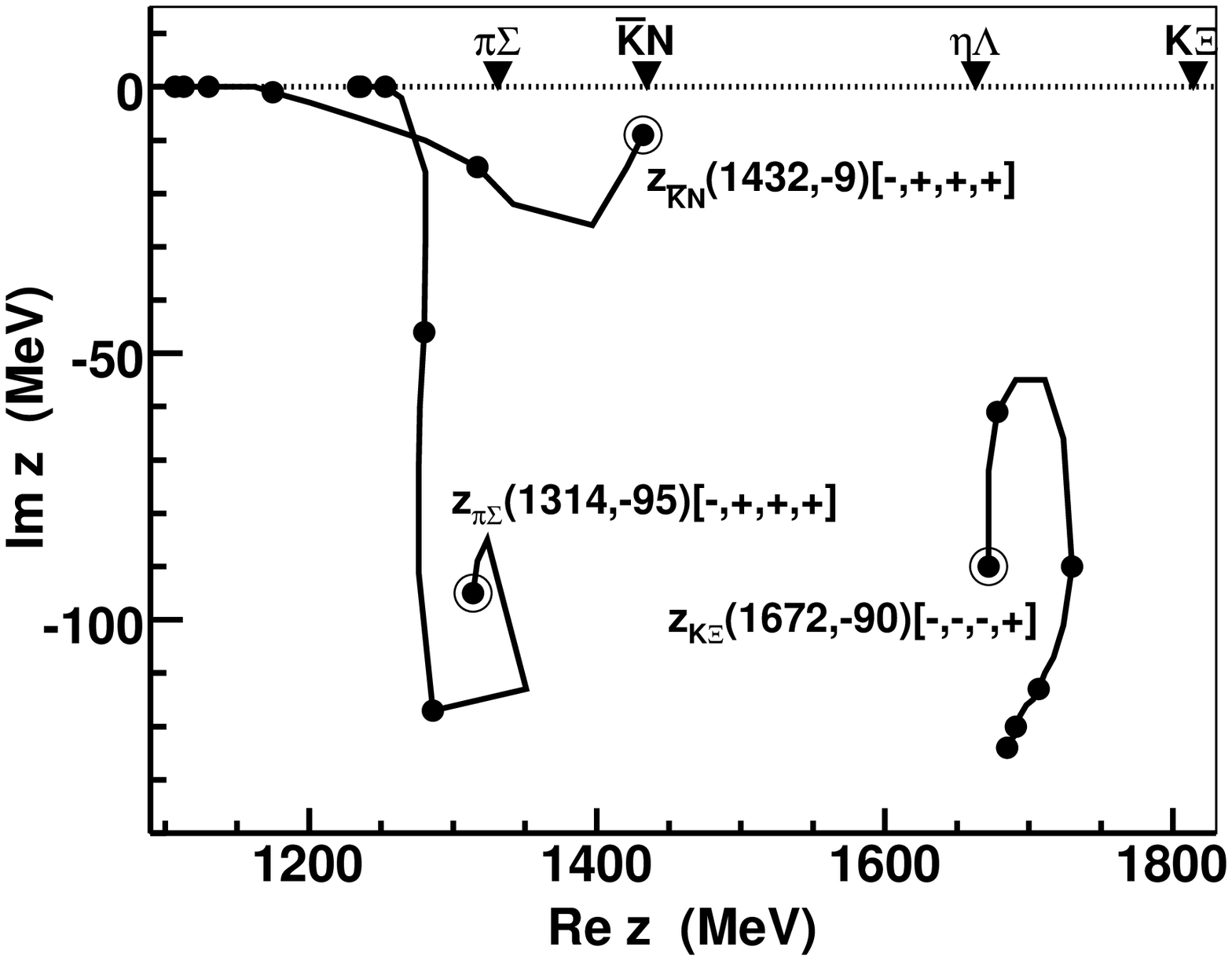}
}
\caption{Trajectories of the isoscalar poles generated for the \MMii{} (left panel) and 
\MMiv{} (right panel) models. For further notation, see Figure \ref{fig:ZCL-CS}.
The $\bar KN$ poles of \MMii{} (\MMiv{}) lie slightly below (above) the $\bar KN$ threshold in the isospin limit.
}

\label{fig:ZCL-MM}       
\end{figure}

Concerning the isoscalar poles, we were able to clearly establish the origin of the two 
poles generated  by the models and assigned to the $\Lambda(1405)$ resonance. As we demonstrate 
in Figure~\ref{fig:ZCL-MM}, 
the dynamics of the \MMii{} model is similar to the \GOii{} one forming the lower 
and higher $\Lambda(1405)$ poles from the $\pi\Sigma$ resonance and the $\eta\Lambda$ 
bound states generated in the ZCL, respectively. On the other hand, in the \MMiv{} model 
the same two poles observed in a physical limit at similar positions trace their origins  
to the $\pi\Sigma$ and $\bar{K}N$ virtual and bound states found in the ZCL, respectively. 
As one can see the \MMiv{} model can also account for the $\Lambda(1670)$ resonance that 
originates from the $K\Xi$ bound state. However, the same pole is missing in the \MMii{} model. 
For this reason as well as because the \MMii{} model generates the higher $\Lambda(1405)$ pole 
from the $\eta\Lambda$ bound state we find the \MMiv{} model more realistic.

Although the pole content of the Bonn models is not so rich as we found for the Murcia models, 
we located several additional isovector poles in the physical limit. Among them only those observed 
for the \MMiv{} model and related to the $K\Xi(I=1)$ bound state in the ZCL had trajectories 
that we managed to follow. When the inter-channel couplings are gradually switched on 
the $K\Xi(I=1)$ pole evolves from its ZCL position on those RSs that share the $+$ sign 
(physical RS) for the $K\Xi$ channel. The shadow poles on the RSs reached directly 
from the physical region by crossing the real axis in between any thresholds end 
their movements (in the physical limit, for $x=1$) at the complex energy 
$z \approx (1720, -20)$ MeV. For the $[-,-,-,+,+]$ RS this pole can be assigned 
to the $\Sigma(1750)$ resonance.

\begin{table}[h]
\caption{The pole positions in the zero coupling limit for the \MMii{} and \MMiv{} models.
See the caption of Table \ref{tab:ZCL-WT} and the text for more details.}
\begin{center}
\begin{tabular}{|cc|lc|lc|}
\hline
         &       & \multicolumn{2}{c|}{\MMii{} model} & \multicolumn{2}{c|}{\MMiv{} model} \\
\hline
 isospin & channel   & $z$(MeV)[$+/-$] &  status & $z$(MeV)[$+/-$] &  status \\ 
\hline
   & $\pi\Sigma$   &  (1366,  $-$77)$[-]$ & R  &  (1324,    0)$[-]$ & V   \\
 0 & $\bar{K}N $   &  $z << z_{\pi\Sigma}({\rm th})$ $[+]$ & B  &  (1107,    0)$[+]$ & B   \\
   & $\eta\Lambda$ &  (1649,    0)$[+]$   & B  &  \multicolumn{1}{c}{---} & -   \\
   & $K\Xi$        &  \multicolumn{1}{c}{---} & -  &  (1685, $-$124)$[+]$ & R   \\
\hline
 1 & $K\Xi$        &  \multicolumn{1}{c}{---} & -  &  (1698,    0)$[+]$ & B   \\
\hline
\end{tabular}
\end{center}
\label{tab:ZCL-MM}
\end{table}

\section{Summary}
\label{sec:summary}

We have presented a comparative analysis of the theoretical approaches that aim at description 
of low-energy meson-baryon interactions in the strangeness $S=-1$ sector. All the models 
discussed in our paper are derived from a chiral Lagrangian that includes terms up to 
the next-to-leading order, ${\cal O}(q^{2})$, in the external meson momenta, with the free parameters
fitted to the low-energy $K^{-}p$ reactions data and to the characteristics of the kaonic hydrogen 
as measured recently by the SIDDHARTA collaboration. As far as we know, this is the first time 
that the various models available on the market were put under a direct comparison aiming 
at determining the subthreshold energy dependence of the $\bar{K}N$ scattering amplitudes 
and on the pole content of the models related to the dynamically generated baryon resonances.

The discussed approaches represent a variety of different philosophies they are built on. 
Most of them (the Kyoto-Munich, Murcia, Bonn ones) use  dimensional regularization to tame the ultraviolet 
divergences in the meson-baryon  loop function and treat the meson-baryon interactions on 
the energy shell while the Prague model introduces off-shell form factors to regularize the Green 
function and phenomenologically accounts for the off-shell effects.
All approaches but the Bonn one are based on a potential concept, introducing 
an effective meson-baryon potential that matches the chiral amplitude up to a given order 
and is then used as a potential kernel in the Lippmann-Schwinger equation to sum a major 
part of the chiral perturbation series. The Bonn model differs by solving a genuine Bethe-Salpeter 
equation before making a projection to the $s$-wave and neglecting the off-shell contributions. 
Finally, the Kyoto-Munich and Prague models have relatively small NLO contributions (representing only moderate 
corrections to the LO chiral interactions) while the Murcia and Bonn models introduce sizable NLO 
terms that generate inter-channel couplings very different from those obtained by only 
the WT interaction. 
Despite all these differences the models are able to reproduce the experimental data on 
a qualitatively very similar level and in mutual agreement especially concerning the data 
available at the $\bar{K}N$ threshold. The models also tend to agree on a position of the 
higher energy of the two poles generated for the $\Lambda(1405)$ resonance, predicting it at 
the complex energy with the real part {$\Re z \approx 1420\ldots1430$~MeV} and the imaginary part 
{$-\Im z \approx 10\ldots40$~MeV}. However, that is  about where the agreement among the model 
predictions ends.

We have demonstrated that the theoretical models considered in our work lead to very different 
predictions for the elastic $K^{-}p$ and $K^{-}n$ amplitudes at sub-threshold energies. There, 
the uncertainty bands (or zones) sketched by the authors of the models in their original 
Refs.~\cite{Ikeda:2012au, Guo:2012vv, Mai:2014xna} represent limits within a setting 
of a particular model rather then general constraints that emerge when the theoretical 
predictions of all models are considered. The latter appear to be much larger then previously 
anticipated. We would like to note here that the predictions made for kaonic atoms and antikaon 
quasi-bound states that were based on in-medium sub-threshold $\bar{K}N$ amplitudes generated 
by the Prague and Kyoto-Munich models and discussed in Refs.~\cite{Cieply:2011yz, Cieply:2011fy, Gazda:2012zz, Friedman:2012qy} 
should be re-examined if the predictions made by the Murcia and Bonn models turn out to be more realistic. 
Indeed, the \GOii{} model and both Bonn models predict much smaller (if any) $\bar{K}N$ attraction 
at subthreshold energies while an in-medium attraction much stronger than provided by any 
of the models considered here is anticipated from phenomenological analysis of 
kaonic atoms \cite{Friedman:1994hx}.

One of the novelties of our paper consists in a detailed analysis of the origins of the poles 
generated in the different approaches. We have done this by following the movement of the poles 
from their {\it physical positions} to the ZCL with inter-channel couplings switched off.
As far as the NLO couplings are small and represent only corrections to the leading order 
WT interaction the pole content of the models is restricted to those poles that 
originate from the $\pi\Sigma$, $\bar{K}N$ and $K\Xi$ channels. This is demonstrated 
by the Prague and Kyoto-Munich models. On the other hand, the Murcia and Bonn models 
discussed here incorporate quite sizable NLO contributions to meson-baryon interactions 
that lead to couplings sufficiently strong to generate poles in the ZCL even in other channels. 
We have found that for two of the NLO models (\GOii{} and \MMii{}) the origin of the higher 
of the two poles assigned to the $\Lambda(1405)$ resonance can be traced to the $\eta \Lambda$ 
bound state in the ZCL. Although one cannot readily dismiss this possibility, it can be argued 
that the other models are more viable as they do not contradict the commonly accepted picture 
of the $\bar{K}N$ bound state that evolves into a resonance due to a strong coupling to 
the $\pi\Sigma$ channel. 

We also note that some of the models predict an existence of the isovector $\bar{K}N$ 
pole on the $[+,+,-,+,+]$ RS. For the \IHWnlo{} model a similar $\eta \Sigma$ pole 
appears on the $[-,-,-,+,+]$ RS. Since the isovector sector is not restricted much by the current 
experimental data the pole can be brought up to a position not far from the $\bar{K}N$ threshold 
and close to the real axis, where it can affect physical observables. The existence of the pole 
was already reported  in Refs.~\cite{Oller:2000fj, Jido:2003cb, Cieply:2011fy} and might 
be related to the isovector contribution to the $\pi\Sigma$ mass spectra observed 
by the CLAS collaboration \cite{Moriya:2013eb}.

Finally, we mention that the analysis of the ZCL solutions of Eqs.~(\ref{eq:ZCLa}) and (\ref{eq:ZCLb}) 
enabled us to determine limits for the inverse ranges or subtractions constants that the fitted 
parameters should satisfy to generate poles in the considered meson-baryon channels. 
The differences among the various models in generating (or in an absence of) a given resonance 
can often be related to the fulfillment of these conditions. This finding is in line 
with similar constraints for the subtraction constants that were discussed within 
a different context in Ref.~\cite{Hyodo:2008xr}.

\vspace*{5mm}{\bf Acknowledgements:}
We would like to acknowledge help by Z.-H.~Guo and Y.~Ikeda who assisted us with 
reproducing the results calculated with the Murcia and Kyoto-Munich models, respectively. 
M.~M.~wishes also to thank Nuclear Physics Institute in \v{R}e\v{z} for the hospitality 
during some phases of this project. The work of A.~C.~was supported by 
the Grant Agency of the Czech Republic, Grant No.~P203/15/04301S. 
The work of M.~M. and U.-G.~M. was supported in part by the DFG and the NSFC through
funds provided to the Sino-German CRC~110 ``Symmetries and the Emergence of
Structure in QCD'' (NSFC Grant No.~11261130311). The work of U.-G.~M. was also 
supported in part by the Chinese Academy of Sciences (CAS) President's International Fellowship 
Initiative (PIFI) (Grant No. 2015VMA076).





\end{document}